\documentclass[aps,reprint,amssymb,amsfonts,amsmath,prl,superscriptaddress,longbibliography]{revtex4-2}
\usepackage{amsfonts}
\usepackage{textcomp,url}
\usepackage{braket,chngcntr,comment,dsfont}
\usepackage{times}
\usepackage{graphicx}
\usepackage{float}
\usepackage{latexsym,amsmath,amssymb,bm,euscript}
\usepackage{color}
\usepackage{subfigure}
\usepackage{epstopdf}
\usepackage[colorlinks=true,linkcolor=blue,citecolor=blue]{hyperref}
\usepackage{soul}
\usepackage[normalem]{ulem}
\usepackage{mathrsfs}
\usepackage{lipsum}
\usepackage{xspace}
\usepackage{natbib}
\usepackage{etoolbox}
\usepackage{tikz,listings}
\usepackage{booktabs,makecell}
\usepackage{pifont}

\newcommand{\Rmnum}[1]{\expandafter\@slowromancap\romannumeral #1@}

\graphicspath{{./figs/}{}}
\begin{document}

\title{$d$-wave FFLO state and charge-2e supersolidity in
    the $t$-$t'$-$J$ model under Zeeman fields}

\author{Xing-Zhou Qu}
\thanks{These authors contributed equally to this work.}
\affiliation{Kavli Institute for Theoretical Sciences, University of Chinese Academy of Sciences, Beijing 100190, China}
\affiliation{Institute of Theoretical Physics, Chinese Academy of Sciences, Beijing 100190, China}

\author{Dai-Wei Qu}
\thanks{These authors contributed equally to this work.}
\affiliation{Arnold Sommerfeld Center for Theoretical Physics, Center for NanoScience, and Munich Center for Quantum Science and Technology, Ludwig-Maximilians-Universität München, 80333 Munich, Germany}
\affiliation{Institute of Theoretical Physics, Chinese Academy of Sciences, Beijing 100190, China}

\author{Qiaoyi Li}
\affiliation{Institute of Theoretical Physics, Chinese Academy of Sciences, Beijing 100190, China}

\author{Wei Li}
\email{w.li@itp.ac.cn}
\affiliation{Institute of Theoretical Physics, Chinese Academy of Sciences, Beijing 100190, China}

\author{Gang Su}
\email{sugang@itp.ac.cn}
\affiliation{Institute of Theoretical Physics, Chinese Academy of Sciences, Beijing 100190, China}
\affiliation{Kavli Institute for Theoretical Sciences, University of Chinese Academy of Sciences, Beijing 100190, China}

\date{\today}

\begin{abstract}
Unconventional superconductivity under strong Zeeman fields—particularly beyond the 
Pauli paramagnetic limit—remains a central challenge in condensed matter physics.
The exotic Fulde–Ferrell–Larkin–Ovchinnikov (FFLO) state, in particular, remains in need of definitive study within fundamental electronic models.
Here we employ state-of-the-art finite-temperature and ground-state tensor network approaches 
to systematically explore the superconducting (SC) phase diagram of the $t$-$t'$-$J$ model subjected to Zeeman fields.
We find that zero-momentum $d$-wave superconductivity persists until the spin gap closes, coexisting with charge density waves.
A novel $d$-wave FFLO phase emerges under a higher Zeeman field even above the Pauli limit, concomitant with a field-enhanced spin density waves.
We identify these phases, characterized by the simultaneous presence of pairing condensate and density wave orders, as charge-2e supersolids.
Analysis of Matsubara Green's function reveals that the FFLO pairing momentum is locked to the underlying Fermi surface. 
Our results provide microscopic insights into field-induced unconventional pairing mechanisms and reveal 
the long-sought FFLO state in a fundamental correlated electron model, offering a promising route 
for its realization in ultracold atom optical lattice.
\end{abstract}

\maketitle

\textit{Introduction.---}
Since the discovery of cuprate superconductors~\cite{cuprate1986}, 
the investigation of high-temperature superconductivity (HTSC) has emerged as a prominent research 
focus in condensed matter physics. 
Numerical studies of strongly correlated electronic models, such as the Hubbard model~\cite{HubbardModel,HubbardModel2}
and $t$-$J$ model~\cite{ZhangRiceSinglet,tJSu}, are widely recognized as a promising avenue for unraveling the pairing mechanisms 
in high-temperature superconductivity.
Notably, while the pure two-dimensional (2D) Hubbard model with large Coulomb repulsion $U$ fails 
to exhibit superconductivity~\cite{Zheng2017stripe,AbsenceSCHubbard,MingpuQinHubbard2022}, 
extensive studies have established that the inclusion of next-nearest-neighbor hopping $t'$ is a 
critical factor enabling $d$-wave superconductivity in the 2D $t$-$J$ model~\cite{Gong2021,JiangHongchen8legtJ,
JiangHongchen2019,JiangHongchen2018-4leg,LE1LE2,White2021,David2022,Plaq_d_wave,LELPhaseStringtJ}. 
In recent years, researchers have also advanced investigations into exotic finite-momentum pairing states, 
such as the pair density wave (PDW) state
\cite{PDWBSCCO,PDWcuprateFujita,PDWcuprateHamidian,PDWcuprateWang,PDWAnnualRev,JiangHongchen3band,ZhangPDW2023,PDWpeps}, 
which is hypothesized to act as a precursor to HTSC.

The Fulde-Ferrell-Larkin-Ovchinnikov (FFLO) state~\cite{FF,LO} can be viewed as a particular type of PDW state, 
distinguished by the breaking of time-reversal symmetry under Zeeman fields~\cite{PDWAnnualRev}.
However, the application of a magnetic field poses significant challenges to theoretical analysis~\cite{tJMagSlaveBoson}
and many-body calculations~\cite{tJMagED,tJladderflux,Wietekflux}. 
Prior researches have identified signatures of the PDW state in the ground state of 1D t-J chains~\cite{1DtJMag} 
and two-leg $t$-$J$ ladders~\cite{White2006,ZhangPDW2023}. Nevertheless, 
the potential existence of FFLO state in the $t$-$t'$-$J$ model under a Zeeman magnetic field remains an open question, 
especially in the finite-temperature regime. 
Further work in this area is indeed anticipated to elucidate the debated pairing mechanisms of HTSC
and to deepen our understanding of the intrinsic properties of cuprate superconducting (SC) states~\cite{CuprateMagAnnualRev}.

In general, superconductivity can be suppressed by magnetic fields via two primary mechanisms: 
the orbital effect~\cite{OrbitalEff1,OrbitalEff2} and the Zeeman effect~\cite{ZeemanEff1,MakiTransitionOrder}. 
The orbital effect originates from the coupling of electron motion to the magnetic vector potential.
This interaction induces vortices within the superconductor, disrupts the coherent superconducting state, 
and ultimately suppresses superconductivity.
The Zeeman effect, by contrast, suppresses superconductivity through a distinct pathway: 
it breaks spin-singlet pairing and lifts spin degeneracy. 
Specifically, superconductivity is suppressed when the Pauli paramagnetic energy, 
that is associated with the alignment of electron spins in the magnetic field, exceeds the SC condensate energy. 
The critical magnetic field at which this transition occurs is known as the Pauli limit (or Clogston-Chandrasekhar limit), 
denoted by $h_{\text{P}}$~\cite{PauliLimit,PauliLimitChandrasekhar}.

\begin{figure}[!tbp]
\centering
\includegraphics[angle=0,width=1.0\linewidth]{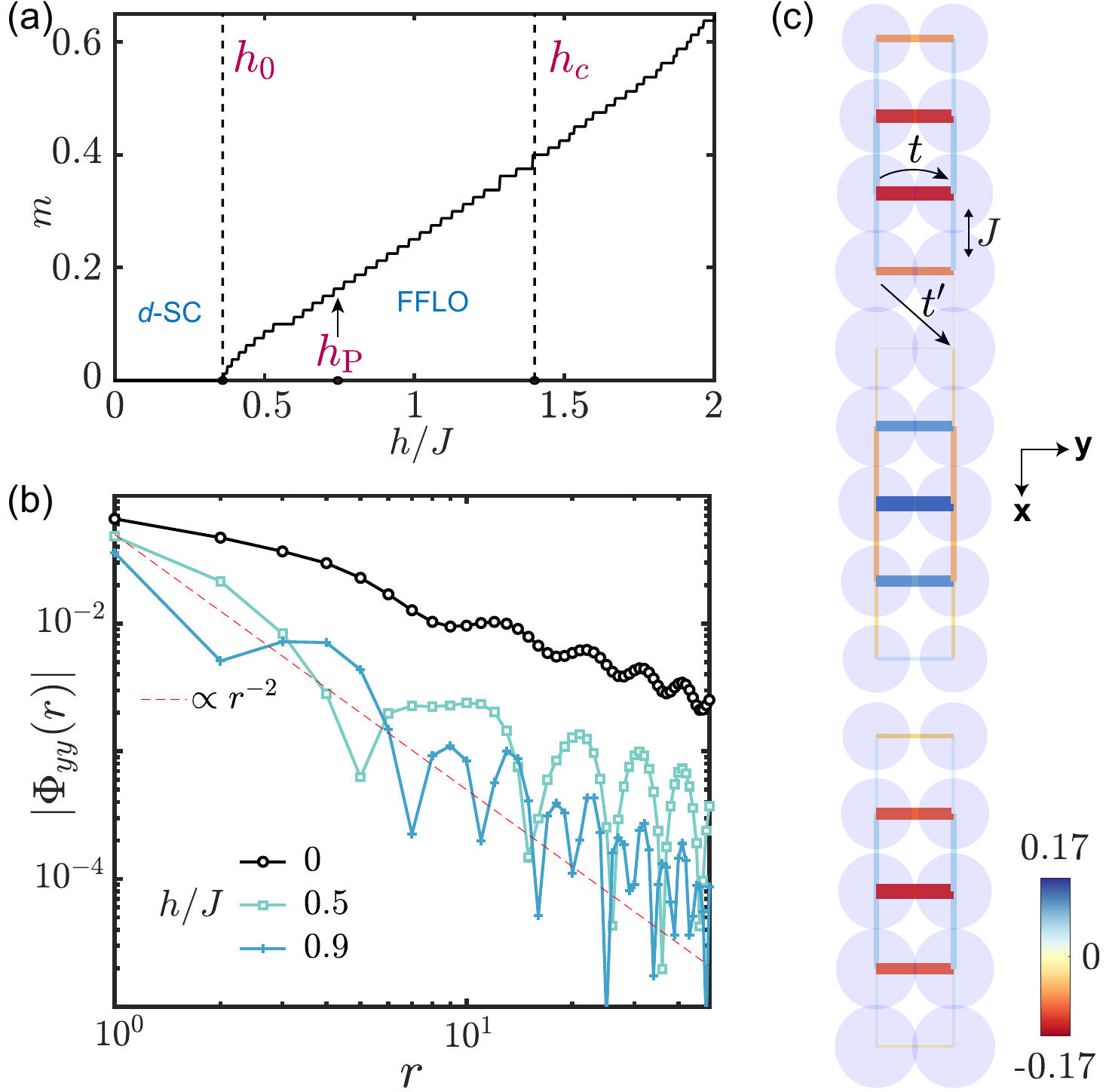}
\caption{
(a) Magnetization $m \equiv (N_\uparrow - N_\downarrow) / N_{\mathrm{site}}$ as a function of Zeeman field $h$ at temperature $T = 0$. 
Here $h_0 = 0.360J$ is the Zeeman field where spin gap closes, 
$h_\mathrm{P} = 0.744J$ denotes the Pauli limit and $h_c \approx 1.4 J$ is the superconducting critical field.
(b) Ground-state pairing correlations $|\Phi_{yy}(r)|$ under different Zeeman fields.
The red dashed line decaying as $r^{-2}$ is also plotted as a reference. $h = 0.5J$ and $h = 0.9J$ locate in the FFLO phase.
(c) Calculated FFLO pairing pattern at $h=0.9J$, 
where the size of circles is proportional to the local hole density, 
and the thickness of bonds represents the pairing amplitude.
Doping $\delta = 0.1$ is fixed in all the calculations.
}
\label{Fig1}
\end{figure}


A well-documented scenario in which SC persists above the Pauli limit is the FFLO state~\cite{FF,LO}. 
Under a strong Zeeman field, FFLO superconductors can accommodate the split Fermi surface (FS) by forming pairs 
with finite center-of-mass momentum, described as $(\mathbf{k}_\uparrow, -\mathbf{k}_\downarrow+\mathbf{q})$, 
where $\mathbf{q}$ denotes the nesting vector between the spin-up and spin-down FS.
In contrast to this straightforward theoretical framework, experimental investigations face substantial 
practical challenges. 
For most materials, the orbital effect is more prominent than the Zeeman effect~\cite{cryst8070285,Makibound,Maki1966}, 
a condition that detrimentally impacts the stability of the FFLO state~\cite{Makibound}. 
To date, the FFLO state has been reported in several classes of materials, 
including quasi-two-dimensional (quasi-2D) organic compounds~\cite{OrganicAndreev,OrganicCalori,OrganicSpecificHeat}, 
iron-based superconductors~\cite{FFLOKFe2As2}, and newly discovered orbital FFLO materials~\cite{Wan2023orbitalFFLO, Itahashi2025Misfit}. 
However, whether the FFLO state can be observed in cuprate superconductors remains elusive.

The present study delves into the interplay between Zeeman field and the SC order in $t$-$t'$-$J$ ladder.
By analyzing pairing correlations and Fermi surface topology, 
we demonstrate the emergence of a FFLO phase and extract the SC onset 
temperature $T_c^*$ via the pairing susceptibility.
We further identify the coexistence of zero-momentum $d$-wave superconductivity + charge density wave (CDW) and the FFLO 
state + spin/charge density wave (SDW/CDW) as charge-2e supersolid (2e-SS) phases,
because there appears the coexistence of density modulation (exhibiting diagonal long-range order, DLRO)
and superconductivity (with off-diagonal long-range order, ODLRO). 
By performing state-of-the-art ground state and finite-temperature tensor network states calculations,
our study establishes the temperature-Zeeman field phase diagram for the $t$-$t'$-$J$ ladder, 
reveals the exotic superconducting phases, and investigates the intertwined charge and spin orders. 
Additionally, this work sheds light on the fundamental connections between superconductivity and magnetism.

\textit{Model and Methods.---}
We investigate the $t$-$t'$-$J$ model under Zeeman fields. The Hamiltonian is given by:
\begin{eqnarray}
\hat{\mathcal{H}} = &-& \sum_{i,j,\sigma} t_{ij} (\hat{c}_{i \sigma}^\dagger \hat{c}_{j \sigma} + h.c.) +
\sum_{i,j} J_{ij} (\hat{\mathbf{S}}_i \cdot \hat{\mathbf{S}}_j - \frac{1}{4}\hat{n}_i \hat{n}_j) \nonumber \\
&-& h \sum_{i} \hat{S}_i^z,
\label{Hamiltonian}
\end{eqnarray}
where $\hat{c}_{i\sigma}$ ($\hat{c}_{i\sigma}^\dagger$) is the electron annihilation (creation) operator with
spin $\sigma = \uparrow, \downarrow$, $\hat{\mathbf{S}}_i$ and $\hat{n}_i$ are spin and density operator at site $i$,
respectively.
We consider the nearest-neighbor and next-nearest-neighbor hoppings ($t$ and $t'$) and antiferromagnetic
exchange interaction ($J$). $h$ denotes the Zeeman field, where the
Land\'e factor $g$ and Bohr magneton $\mu_\text{B}$ are absorbed.
The double occupancy on a site is projected out.
We choose $J=1$, $t/J=3$ and $t'/t= 0.17$ in following calculations. 

Our numerical study employs both zero-temperature DMRG~\cite{DMRGprl,DMRGmps} and finite-temperature 
tanTRG~\cite{tanTRG2023,Li_FiniteMPS_jl_2025} methods, respecting $\mathrm{U(1)}_{\text{charge}} \times \mathrm{U(1)}_{\text{spin}}$ symmetry,
which is implemented by the tensor network library QSpace~\cite{QSpace} and TensorKit.jl~\cite{TensorKit}.
For the main text, we study the model on a ladder whose length $L_x$ is up to $80$; corresponding results for a
width-4 cylinder are provided in the Supplemental Material (SM)~\cite{SM}.
For ground-state DMRG calculations we keep 3000 $\mathrm{U(1)}_{\text{charge}} \times \mathrm{U(1)}_{\text{spin}}$
states, ensuring a typical truncation error $\epsilon < 1 \times 10^{-7}$. 
In finite-temperature calculations, we retain 2800 states to ensure robust convergence~\cite{SM} at system temperatures as low as $T = 0.02J$.
The hole doping is fixed at $\delta = 0.1$
in all calculations, through specifing the charge quantum number in $T=0$ DMRG and fine 
tuning the chemical potential~\cite{li2025fixNtanTRG} in $T>0$ tanTRG methods.

\textit{Magnetization Curve and the $d$-wave FFLO superconductivity.---}
We begin by examining the ground-state properties of the $t$-$t'$-$J$ ladder under a Zeeman field. 
The magnetization curve in Fig.~\ref{Fig1}(a) is obtained through DMRG calculations, from which
we acquire the critical field $h_0 = 0.36 J$ where spin gap closes.
It is seen that the system exhibits a robust zero-momentum $d$-wave superconductivity ($d$-SC) below $h_0$,
consistent with previous studies under specified parameters in the absence of magnetic field~\cite{Gong2021,LE1LE2,LuXintwoleg}.
The Pauli limit is estimated by the ground-state energy among the states with different
quantum numbers~\cite{White2006} 
$h_{\textrm{P}} = [E(n_{\textrm{h}}-1, S^z +1/2) + E(n_{\textrm{h}}-1, S^z -1/2)
- E(n_{\textrm{h}}, S^z) - E(n_{\textrm{h}}-2, S^z)]\mid_{S^z =0}$, where $E(n_{\textrm{h}}, S^z)$
denotes the ground-state energy with hole number $n_{\textrm{h}}$ and spin quantum number $S^z$.
Accordingly we obtain the Pauli limit $h_{\textrm{P}} = 0.744 J$ for our system as indicated in Fig.~\ref{Fig1}(a).
The slope near $h \approx 1.4 J$ indicates the SC critical field, above which the system 
loses superconductivity and evolves into a Luttinger liquid phase.
Moreover, we notice the magnetization plateau at $m = \delta = 0.1$, corresponds to
the existence of spin gap near $h = 0.6 J$, which satisfies the OYA relation~\cite{OYA,OYA1D}.

To explore the SC order, we calculate the rung pairing correlations
$\Phi_{yy}(\mathbf{r}_i,\mathbf{r}_j) = \langle \hat{\Delta}_y^\dagger(\mathbf{r}_i) \hat{\Delta}_y(\mathbf{r}_j) \rangle$,
where $\hat{\Delta}_y (\mathbf{r}_i) = (\hat{c}_{i,\uparrow}\hat{c}_{i+\hat{y},\downarrow}
- \hat{c}_{i,\downarrow} \hat{c}_{i+\hat{y},\uparrow})/\sqrt{2}$ is the rung singlet pairing operator at site $i$.
In Fig.~\ref{Fig1}(b), we find that the SC correlations $|\Phi_{yy}(r)|$ ($r = |\mathbf{r}_i - \mathbf{r}_j|$) are suppressed by the Zeeman field, 
displaying smaller amplitudes and a more rapid decay. 
Nevertheless, $|\Phi_{yy}(r)|$ retains a quasi-long-range order and decays slower than $r^{-2}$ ($|\Phi_{yy}(r)| \propto r^{-1.55}$ at $h/J = 0.9$),
which implies a robust SC quasi-long range order. 
Note that in the FFLO phase, spin correlations exhibit a power-law decay~\cite{SM}. 
This behavior contrasts with the Luther-Emery liquid at $h=0$, and general expressions for the Luttinger parameters remain to be derived~\cite{Pruschke1992superconducting}.

In Fig.~\ref{Fig1}(c), we visualize the $d$-wave FFLO pairing pattern in real space~\cite{Wietek2022,Wietekflux}.
Given the DMRG ground state, we calculate the singlet pairing correlation between all bonds and construct the two-particle reduced density matrix
$\rho_S (\mathbf{r}_i, \alpha | \mathbf{r}_j, \alpha') = \langle \hat{\Delta}_\alpha^\dagger (\mathbf{r}_i) \hat{\Delta}_{\alpha'} (\mathbf{r}_j) \rangle$, 
where $\alpha, \alpha' = \hat{x}, \hat{y}$~is the unit vector. 
Diagonalizing $\rho_S$ yields dominant eigenvalues, and the associated eigenvectors encode both the magnitude and the symmetry of the SC pairing. 
In Fig.~\ref{Fig1}(c) we extract the central region of the system and display the leading pairing pattern 
at $T = 0$ and $h = 0.9J$. The pattern reveals a spatial modulation of superconducting stripes with sign changes 
between the $x$- and $y$-bonds, confirming the emergence of a $d$-wave FFLO state.

\begin{figure}[tbp]
\centering
\includegraphics[angle=0,width=\linewidth]{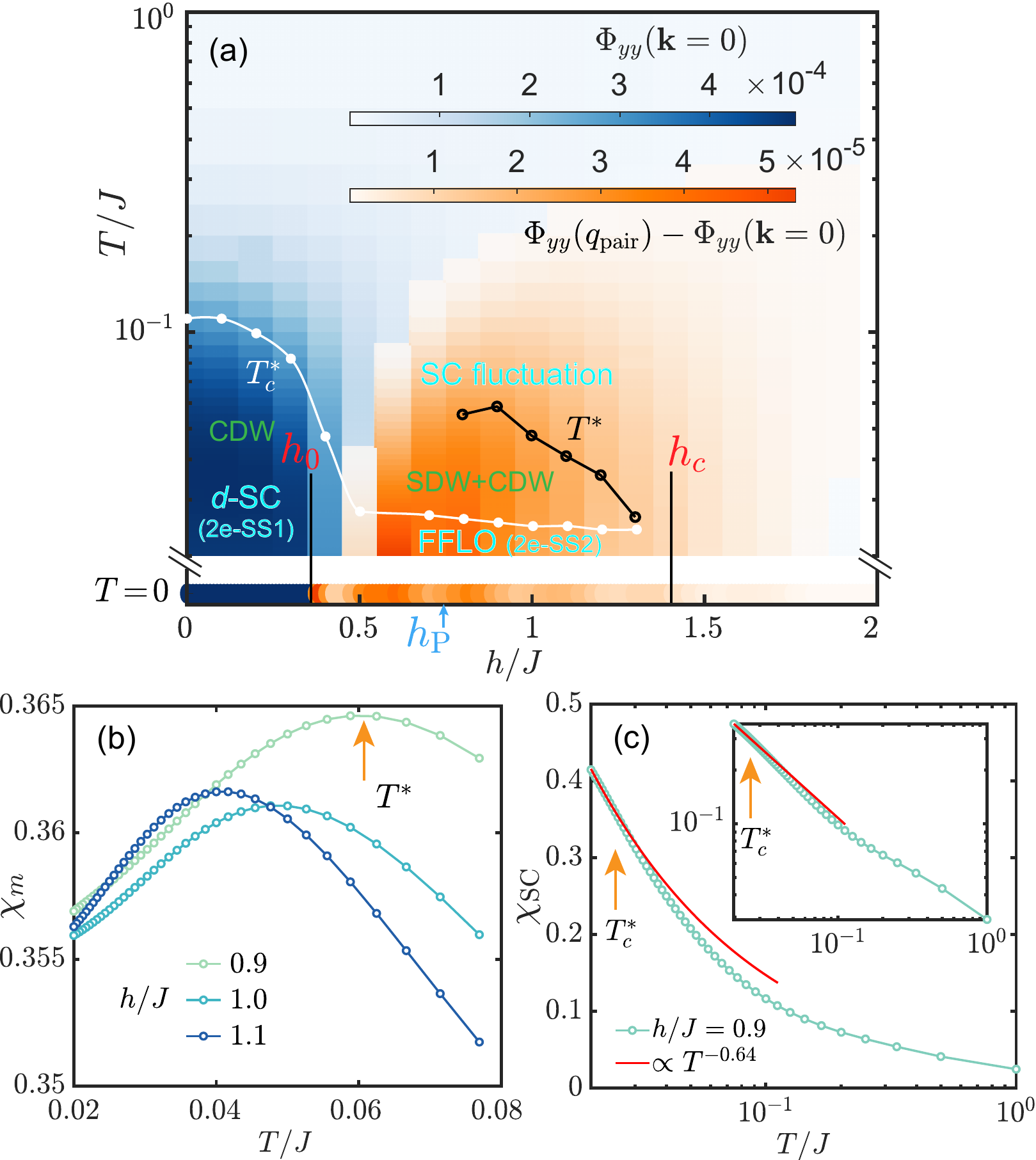}
\caption{(a) Finite-temperature and ground-state phase diagram of $t$-$t'$-$J$ ladder ($t'/t > 0$) in the presence of a Zeeman field $h$. 
For Zeeman fields below $h_0$, we identify a zero-momentum $d$-wave superconducting phase ($d$-SC) emerging at $T_c^*$ that coexists with a 
charge density wave, which constitutes a charge-2e supersolid (2e-SS1).
The magnitude of $\Phi_{yy}(\mathbf{k=0})$ (indicated by the blue colorbar) reflects the strength of the zero-momentum pairing correlations.
For $h > h_0$, the $d$-SC persists at elevated temperatures.
The finite-momentum superconducting phase emerges with intertwined spin density wave and charge density wave, 
which constitutes another charge-2e supersolid phase (2e-SS2).
Here $T_c^*$ denotes the onset temperatures of $d$-wave FFLO superconductivity.
The difference $\Phi_{yy}(\mathbf{q}_{\mathrm{pair}}) - \Phi_{yy}(\mathbf{k=0})$ (depicted by the orange colorbar) 
quantifies the relative magnitude of the finite-momentum pairing correlations, 
also capturing the finite-momentum SC fluctuations at high temperatures.
(b) Magnetic susceptibility $\chi_m = \partial m / \partial h$ in the FFLO (fluctuating) phase. 
$T^*$ denotes the magnetic characteristic temperature at which spin density wave emerges.
(c) Pairing susceptibility $\chi_{\mathrm{SC}} = \partial \langle \Delta_{\mathrm{tot}} \rangle / \partial h_{\mathrm{pair}}$
in the FFLO (fluctuating) phase, which diverges algebraically as $\chi_{\mathrm{SC}} \sim 1/T^\gamma$ below the SC onset temperature $T_c^*$. 
The red line corresponds to the fitted power-law curve.
}
\setlength{\dbltextfloatsep}{1pt plus 1.0pt minus 2.0pt}
\setlength{\belowdisplayskip}{3pt}
\label{Fig2}
\end{figure}

\begin{figure}[tbp]
\centering
\includegraphics[angle=0,width=1.0\linewidth]{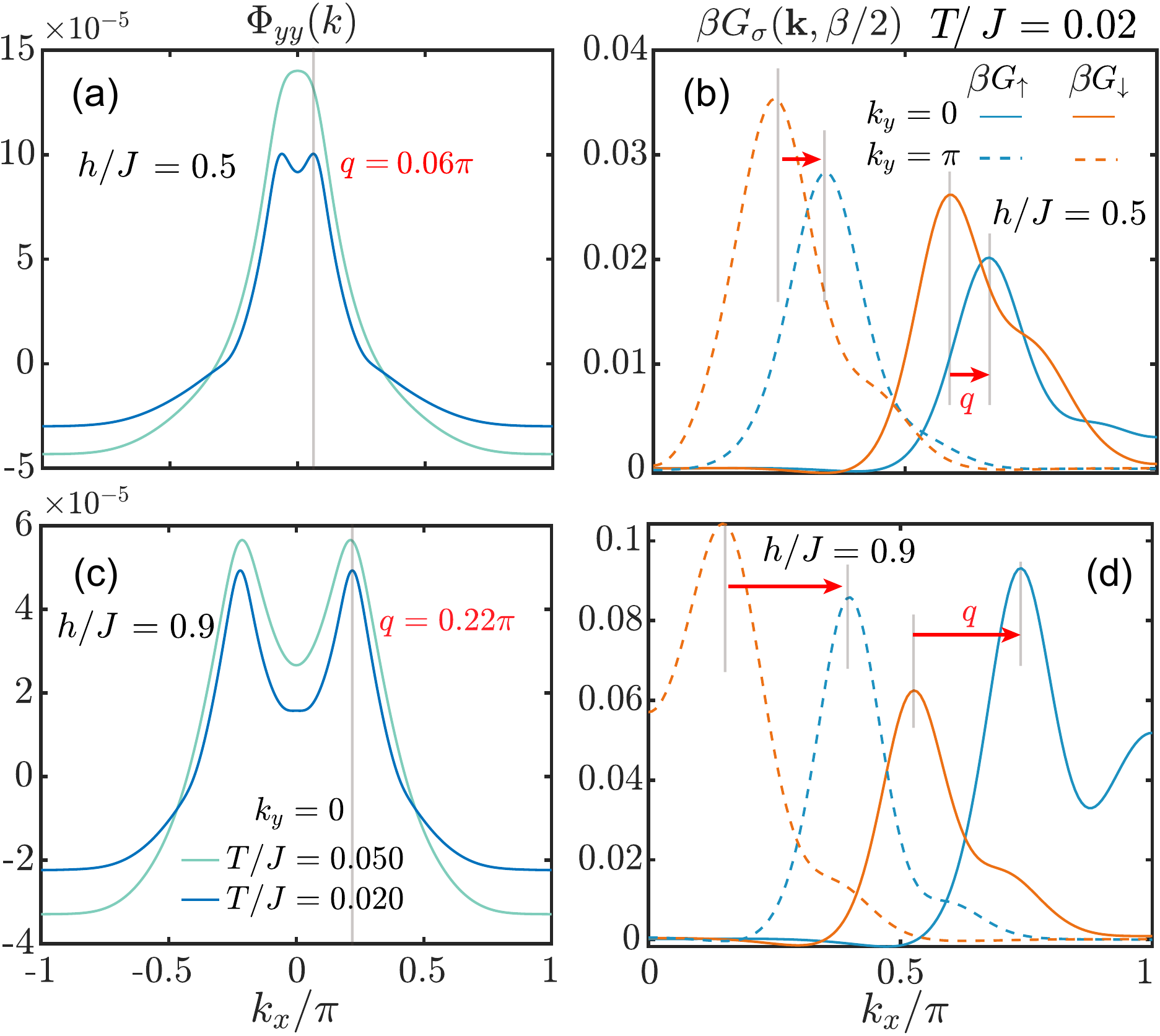}
\caption{(a,c) Rung pairing correlation $\Phi_{yy}(k_x, k_y=0)$ at $T = 0.02 J$ and $T = 0.05 J$.
(b, d) $\beta G_\sigma (\mathbf{k}, \beta/2)$ at low temperature $\beta = 50 J$ ($T = 0.02 J$). 
In (a) and (b) we take $h = 0.5 J$, exhibiting zero-momentum $d$-wave superconductivity at high temperature 
while FFLO state appearing at low temperature.
(c, d) show the FFLO state with finite-momentum pairing ($h = 0.9 J$).}
\setlength{\dbltextfloatsep}{1pt plus 1.0pt minus 2.0pt}
\setlength{\belowdisplayskip}{3pt}
\label{Fig3}
\end{figure}

\begin{figure}[tbp]
\centering
\includegraphics[width=0.5\textwidth]{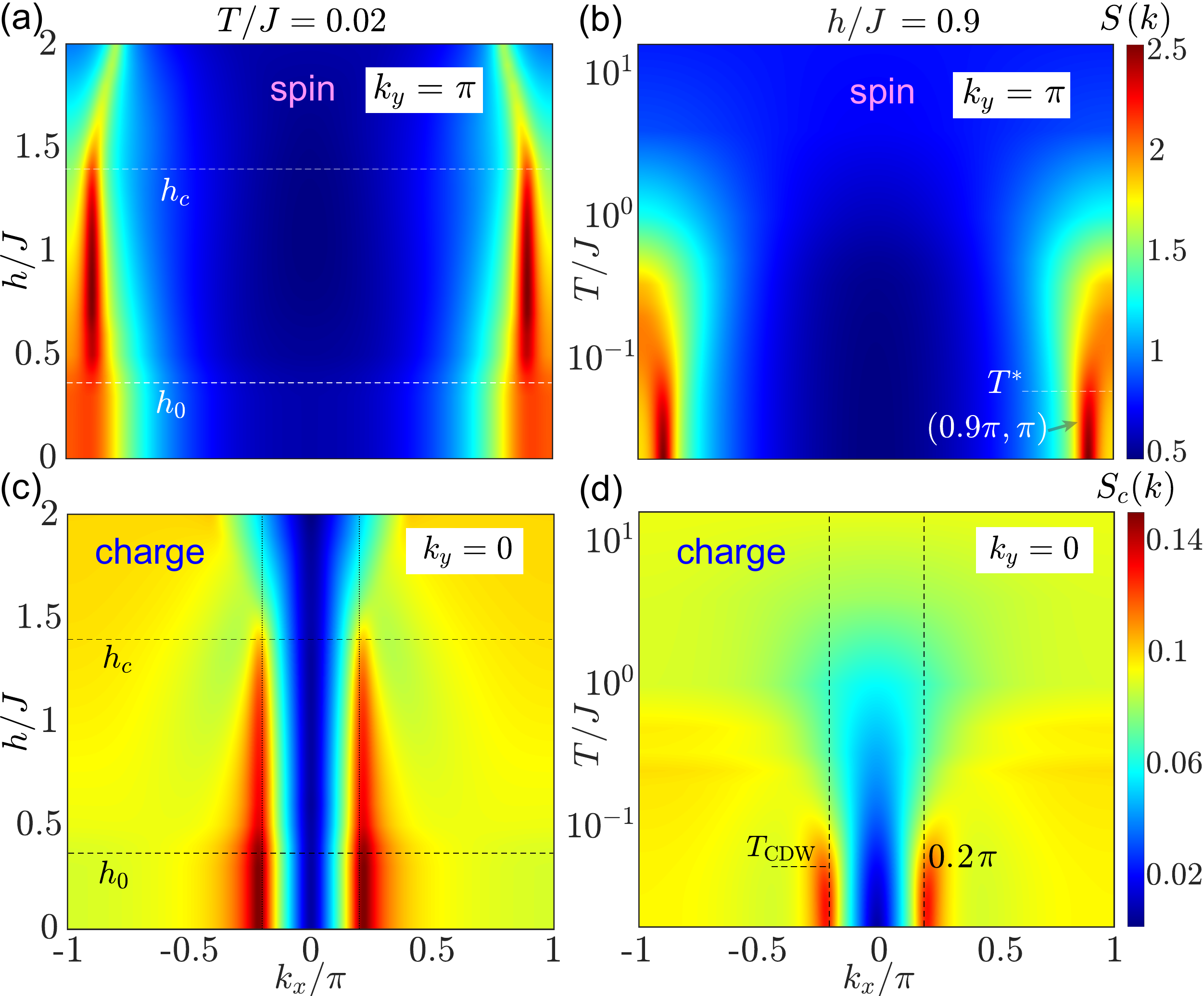}
\caption{Contour plots of the spin structure factor $S(k_x, k_y = \pi)$ and charge structure factor 
$S_c (k_x, k_y = 0)$. In (a, c), the temperature is fixed at $T = 0.02 J$. Both the spin gapless point
$h_0$ and the superconducting critical field $h_{c}$ are identified.
In (b, d) the Zeeman field is set at $h = 0.9 J$ (2e-SS2 phase).
The magnetic characteristic temperatures $T^*$ is indicated.
}
\label{Fig4}
\end{figure}

\textit{Finite-temperature phase diagram, magnetic and pairing susceptibility.---}
We then study the finite-temperature properties by focusing 
on the SC pairing correlations.
To distinguish the finite-momentum pairing from the zero-momentum $d$-wave pairing,
we calculate the Fourier transform of rung pairing correlations
$\Phi_{yy}(\mathbf{k}) = 1/N \sum_{i, j}
\mathrm{e}^{\mathrm{i} \mathbf{k} \cdot (\mathbf{r}_i - \mathbf{r}_j)} \Phi_{yy}(\mathbf{r}_i,\mathbf{r}_j)$.
Fig.~\ref{Fig2}(a) shows the pairing correlations, with blue contours indicating the magnitude of 
zero-momentum pairing $\Phi_{yy}(\mathbf{k=0})$.
In the $h < h_0$ regime, the superconductivity has a zero-momentum $d$-wave pairing, corresponding to the $S^z = 0$ ground state 
reflected by the zero-magnetization plateau in Fig.~\ref{Fig1}(a).
The role of Zeeman field is to slightly suppress the SC temperature.
As the Zeeman field increases above $h_0$, a finite-momentum $\mathbf{q}_{\mathrm{pair}} \neq 0$ pairing mode emerges 
in the low temperature regime (see also Fig.~\ref{Fig3}(a)) while the $d$-SC survives 
at high temperature until the Pauli limit $h_{\mathrm{P}}$. 
The appearance of finite-momentum pairing fluctuation is highlighted by the difference $\Phi_{yy}(\mathbf{q}_{\mathrm{pair}}) - \Phi_{yy}(\mathbf{k=0})$
in orange contours.
Above the Pauli limit ($h > h_\mathrm{P}$), SC fluctuations emerge in conjunction with the 
breaking of translational symmetry. Here the zero-momentum $d$-wave pairing is entirely suppressed,
while the FFLO fluctuation is prominent at high temperatures and the dominant FFLO superconducting phase develops below $T_c^*$ 
(see also Fig.~\ref{Fig3}(c)).
We observe the coexisting charge orders in the $d$-SC phase (cf. Fig.~\ref{Fig4}(c)) and the field enhanced 
SDW in the FFLO phase (cf. Fig.~\ref{Fig4}(a)). 
In these regimes, the holes form Cooper pairs that condensate and move without resistance (superfluidity) coexisting with the
charge or spin density modulations that spontaneously breaks the translational symmetry (solidity), 
giving rise to the macroscopic quantum states coined as the charge-2e supersolid states (2e-SS1 and 2e-SS2)~\cite{Leggett1970SS,Xiang2024SS,Recati2023SS,xie2025SS,Meng2025SS},
as these states possess simultaneously both DLRO and ODLRO.

In the FFLO phase, we calculate the magnetic susceptibility $\chi_{m} = \partial m / \partial h$, as 
shown in Fig.~\ref{Fig2}(b).
In the absence of magnetic field, the previous study~\cite{David2022} shows that $\chi_m$ reaches a 
maximum and then vanishes, which indicates the onset of pseudogap~\cite{pseudogap2001,pseudogap2013,pseudogap2009}.
Here we uncover within the FFLO phase, $\chi_m$ is suppressed due to the formation of spin orders,
whose maximum can be extracted and employed as a characteristic temperature $T^*$.
Moreover, the Zeeman field in this instance produces a spin-gapless ground state~\cite{SM},
making $\chi_m$ converges to a finite value as $T \rightarrow 0$.
We also study the SC pairing susceptibility 
$\chi_{\mathrm{SC}} = \partial \langle \Delta_{\mathrm{tot}} \rangle / \partial h_{\mathrm{pair}}$, 
where we introduce a local pairing field $h_{\mathrm{pair}}$ coupled to a rung singlet and calculate the induced 
$\Delta_{\mathrm{tot}} = \frac{1}{2} \sum_{\mathbf{r}, \alpha} |\Delta_\alpha (\mathbf{r}) + \Delta_\alpha^\dagger (\mathbf{r})|$.
$\chi_{\mathrm{SC}}$ exhibits an algebraic divergence behavior of $\chi_{\mathrm{SC}} \sim 1 / T^\gamma$ below the 
onset temperature $T_c^*$, with $\gamma \approx 0.64$ at $h /J = 0.9$, consistent with the SC quasi-long-range order confirmed 
in the ground-state.
The fitted $T_c^*$ is plotted by the white lines in Fig.~\ref{Fig2}(a).
The relative low $T_c^*$ in the 2e-SS2 phase indicates that
the FFLO superconductivity emerges at a considerably lower temperature scale compared to the zero-momentum $d$-wave superconducting 
phase owing to the effect of Zeeman fields.

\textit{Finite-momentum pairing and Fermi surface locking.---}
To analyze the relationship between finite-momentum pairing and FS, we calculate the Matsubara Green’s 
function $G_{\sigma}(\mathbf{k}, \beta / 2) = \langle e^{\beta \hat{\mathcal{H}} / 2}
\hat{c}_{\mathbf{k}, \sigma}^\dagger e^{-\beta \hat{\mathcal{H}} / 2} \hat{c}_{\mathbf{k}, \sigma} \rangle_\beta$,
where $\hat{c}_{\mathbf{k}, \sigma} = 1/\sqrt{N} \sum_{i} e^{-\mathrm{i}\mathbf{k}\cdot \mathbf{r}_i}\hat{c}_{i, \sigma}$
and $\beta = 1/ k_\mathrm{B}T$ is the inverse temperature.
In the low-temperature limit $\beta \gg 1$, one can approximate the spectral weight around Fermi level
$\beta G_{\sigma}(\mathbf{k}, \beta / 2) \sim A_\sigma (\mathbf{k}, \omega = 0)$ \cite{Jiang2022ITP,Samuel2017ITP}.
In Fig.~\ref{Fig3} $\Phi_{yy}(k_x, k_y = 0)$ and $\beta G_{\sigma}(\mathbf{k}, \beta / 2)$ are shown at different temperatures.
At $h = 0.5 J$, $\Phi_{yy}(\mathbf{k})$ in Fig.~\ref{Fig3}(a) 
peaks at $k_x = 0$ at high temperature but develops a finite-momentum $q = 0.06 \pi$ pairing order at lower temperatures.
In this case, the zero-momentum $d$-wave pairing order remains energetically favorable at high temperatures,
whereas the $\mathbf{q} \neq 0$ pairing order dominants only at low temperatures.
Figs.~\ref{Fig3}(b,d) present the $\beta G_{\sigma}(\mathbf{k}, \beta / 2)$ at low temperature $T = 0.02 J$.
The present system under interest has a FS with two branches $k_y = 0, \pi$. 
The nesting vector (marked by red arrows) can be extracted from the difference between the spin up and spin down components of the FS,
which roughly align with the pairing momentum identified by the peak of $\Phi_{yy}(\mathbf{k})$,
showing the finite-momentum pairing and FS locking and also confirming a conventional FFLO phase.


\textit{Intertwined charge and spin orders.---}
To understand the charge and spin orders in the superconductive phases, 
we show in Fig.~\ref{Fig4} the contour plot of the spin structure factor 
$S(\mathbf{k}) = 1/N \sum_{i, j} e^{\mathrm{i}\mathbf{k}\cdot(\mathbf{r}_i - \mathbf{r}_j)} 
\left( \langle \hat{\mathbf{S}}_i \cdot \hat{\mathbf{S}}_j \rangle - \langle \hat{S}_i^z \rangle \langle \hat{S}_j^z \rangle \right)$
and the charge structure factor 
$S_c(\mathbf{k}) = 1/N \sum_{i, j} e^{\mathrm{i}\mathbf{k}\cdot(\mathbf{r}_i - \mathbf{r}_j)} 
\left( \langle \hat{n}_i \hat{n}_j \rangle - \langle \hat{n}_i \rangle \langle \hat{n}_j \rangle \right)$.
In Fig.~\ref{Fig4}(a), we fix a low temperature $T = 0.02J$ and show the field dependence of $S(k_x, k_y = \pi)$.
The system exhibits pronounced antiferromagnetic fluctuations on the zero magnetization plateau $h < h_0$.
As the Zeeman field increases, a $((1-\delta)\pi, \pi)$ SDW
emerges concomitantly with the finite-momentum pairing and persists until superconductivity vanishes.
In Fig.~\ref{Fig4}(b), at a fixed Zeeman field $h = 0.9J$, we study the temperature evolution of spin orders. 
It can be seen that the antiferromagnetic fluctuation is suppressed as the temperature decreases,
while the $((1-\delta)\pi, \pi)$ SDW becomes dominant below the magnetic characteristic temperature $T^*$.
Figs.~\ref{Fig4}(c,d) manifest the charge structure factor $S_c(k_x, k_y = 0)$ peaks at $(2\delta \pi, 0)$,   
which present the field dependence and temperature evolution of charge order in the FFLO phase.
We further inspect the charge density profiles in the SM~\cite{SM} and ensure a typical ``filled stripe" state~\cite{LE1LE2,LuXintwoleg,SM}.
In contrast to the field-enhanced SDW, the magnitude of CDW is suppressed by the Zeeman field (cf. Fig.~\ref{Fig4}(c)).
Fig.~\ref{Fig4}(d) illustrates that the charge order emerges below the characteristic temperature $T_{\mathrm{CDW}}$, 
which is estimated by the maximum of temperature gradient $\frac{\partial}{\partial T} S_c(T)|_{\mathbf{k} = (2\delta \pi, 0)}$.

Furthermore, we find that both SDW and CDW wave vectors are not affected by the Zeeman field, 
precluding the possibility that they are developed from the FFLO order~\cite{PDWAnnualRev}. 
Instead, the intertwined charge and spin orders have coherent wave length~\cite{ZhangShiweiStripe2022}
that depends solely on doping $\delta$.
It is interesting to note that based on these results, one finds that in the $d$-SC and FFLO phase, 
the charge-2e SC condensate (with ODLRO) can coexist with the CDW and SDW orders (with DLRO), 
showing the existence of supersolid phases.

\textit{Discussion.---}
The interplay between magnetic field and unconventional superconductivity constitutes a fundamental and significant issue.
In this work, we delve into the Zeeman effect in the $t$-$t'$-$J$ model and present a comprehensive 
phase diagram for both ground-state and finite-temperature regimes.
It is found that the robust zero-momentum $d$-wave SC order survives at low Zeeman field and 
a $d$-wave FFLO state exists at even higher Zeeman field, 
which is manifested by the finite-momentum pairing with quasi long-range SC order.
The temperature evolution of the FFLO phase is discussed based on the SC pairing susceptibility.
We further study the effect of the FS, and found that it is split by the Zeeman field and locks to the 
pairing momentum of FFLO state. 
The intertwined spin and charge orders expose an enhanced SDW and suppressed CDW as the emergence of FFLO, 
whose wave vectors are closely related to the doping. 
The coexisting charge/spin density waves with SC orders form a supersolidity in the system under certain characteristic temperatures.

Taking the same model parameters, we also calculate the ground-state properties of four-leg $t$-$t'$-$J$ cylinders 
and include the DMRG results in the SM~\cite{SM}. The results reveal that the FFLO state is also presented 
in the $t' / t > 0$ regime with intertwined charge and spin orders, indicating that
the FFLO superconductivity and supersolidity could be an intrinsic feature in the $t$-$t'$-$J$ model
under a Zeeman field.
Nevertheless, given the spin exchange $J \approx 125$~meV in cuprates~\cite{lyons1988dynamics},
the Pauli limit is estimated to be $h_{\text{P}} = 0.744 J / g\mu_B \approx 800 \mathrm{T}$, which is too large to be readily reached, which 
could help account for the absence of FFLO experimental evidence in cuprates~\cite{LSCOparallfield}.

In addition, our present work provides a guidance for experiments with ultracold atoms. Recently, $s$-wave FFLO pairing state was reported 
in ultracold atom systems modeled by an attractive Hubbard Hamiltonian in the intermediate interaction regime~\cite{feng2025search}. 
Since Zeeman splitting can be readily implemented via spin imbalance~\cite{feng2025search,Kinnunen_2018,Revelle20161Dto3D,Liao2010SpinImbalance}, 
we expect that $d$-wave FFLO state and charge-2e supersolid could likewise be observed in optical lattices, 
enabling direct comparison with our calculated results in upcoming experiments.

\begin{acknowledgments}
    \textit{Acknowledgment.---}
    The authors acknowledge Junsen Wang and Haoxin Wang for helpful discussions.
    This work was supported by NSFC Nos. 12534009 and 12447101, 
    the Quantum Science and Technology-National Science and Technology Major Project under Grant 
    No. 2024ZD0300500, the Strategic Priority Research Program of Chinese Academy of Sciences (Grant No. XDB1270000) 
    and the CAS Superconducting Research Project under Grant No. SCZX-0101.
\end{acknowledgments}

\bibliography{ttpJMag}

\begin{thebibliography}{79}%
\makeatletter
\providecommand \@ifxundefined [1]{%
 \@ifx{#1\undefined}
}%
\providecommand \@ifnum [1]{%
 \ifnum #1\expandafter \@firstoftwo
 \else \expandafter \@secondoftwo
 \fi
}%
\providecommand \@ifx [1]{%
 \ifx #1\expandafter \@firstoftwo
 \else \expandafter \@secondoftwo
 \fi
}%
\providecommand \natexlab [1]{#1}%
\providecommand \enquote  [1]{``#1''}%
\providecommand \bibnamefont  [1]{#1}%
\providecommand \bibfnamefont [1]{#1}%
\providecommand \citenamefont [1]{#1}%
\providecommand \href@noop [0]{\@secondoftwo}%
\providecommand \href [0]{\begingroup \@sanitize@url \@href}%
\providecommand \@href[1]{\@@startlink{#1}\@@href}%
\providecommand \@@href[1]{\endgroup#1\@@endlink}%
\providecommand \@sanitize@url [0]{\catcode `\\12\catcode `\$12\catcode
  `\&12\catcode `\#12\catcode `\^12\catcode `\_12\catcode `\%12\relax}%
\providecommand \@@startlink[1]{}%
\providecommand \@@endlink[0]{}%
\providecommand \url  [0]{\begingroup\@sanitize@url \@url }%
\providecommand \@url [1]{\endgroup\@href {#1}{\urlprefix }}%
\providecommand \urlprefix  [0]{URL }%
\providecommand \Eprint [0]{\href }%
\providecommand \doibase [0]{https://doi.org/}%
\providecommand \selectlanguage [0]{\@gobble}%
\providecommand \bibinfo  [0]{\@secondoftwo}%
\providecommand \bibfield  [0]{\@secondoftwo}%
\providecommand \translation [1]{[#1]}%
\providecommand \BibitemOpen [0]{}%
\providecommand \bibitemStop [0]{}%
\providecommand \bibitemNoStop [0]{.\EOS\space}%
\providecommand \EOS [0]{\spacefactor3000\relax}%
\providecommand \BibitemShut  [1]{\csname bibitem#1\endcsname}%
\let\auto@bib@innerbib\@empty
\bibitem [{\citenamefont {{Bednorz}}\ and\ \citenamefont
  {{M{\"u}ller}}(1986)}]{cuprate1986}%
  \BibitemOpen
  \bibfield  {author} {\bibinfo {author} {\bibfnamefont {J.~G.}\ \bibnamefont
  {{Bednorz}}}\ and\ \bibinfo {author} {\bibfnamefont {K.~A.}\ \bibnamefont
  {{M{\"u}ller}}},\ }\bibfield  {title} {\bibinfo {title} {{Possible high
  T$_{c}$ superconductivity in the Ba-La-Cu-O system}},\ }\href
  {https://doi.org/10.1007/BF01303701} {\bibfield  {journal} {\bibinfo
  {journal} {Zeitschrift fur Physik B Condensed Matter}\ }\textbf {\bibinfo
  {volume} {64}},\ \bibinfo {pages} {189} (\bibinfo {year} {1986})}\BibitemShut
  {NoStop}%
\bibitem [{\citenamefont {Hubbard}(1964)}]{HubbardModel}%
  \BibitemOpen
  \bibfield  {author} {\bibinfo {author} {\bibfnamefont {J.}~\bibnamefont
  {Hubbard}},\ }\bibfield  {title} {\bibinfo {title} {{Electron correlations in
  narrow energy bands. II. The degenerate band case}},\ }\href@noop {}
  {\bibfield  {journal} {\bibinfo  {journal} {Proceedings of the Royal Society
  of London. Series A. Mathematical and Physical Sciences}\ }\textbf {\bibinfo
  {volume} {277}},\ \bibinfo {pages} {237} (\bibinfo {year}
  {1964})}\BibitemShut {NoStop}%
\bibitem [{\citenamefont {Gutzwiller}(1963)}]{HubbardModel2}%
  \BibitemOpen
  \bibfield  {author} {\bibinfo {author} {\bibfnamefont {M.~C.}\ \bibnamefont
  {Gutzwiller}},\ }\bibfield  {title} {\bibinfo {title} {{Effect of Correlation
  on the Ferromagnetism of Transition Metals}},\ }\href
  {https://doi.org/10.1103/PhysRevLett.10.159} {\bibfield  {journal} {\bibinfo
  {journal} {Phys. Rev. Lett.}\ }\textbf {\bibinfo {volume} {10}},\ \bibinfo
  {pages} {159} (\bibinfo {year} {1963})}\BibitemShut {NoStop}%
\bibitem [{\citenamefont {Zhang}\ and\ \citenamefont
  {Rice}(1988)}]{ZhangRiceSinglet}%
  \BibitemOpen
  \bibfield  {author} {\bibinfo {author} {\bibfnamefont {F.~C.}\ \bibnamefont
  {Zhang}}\ and\ \bibinfo {author} {\bibfnamefont {T.~M.}\ \bibnamefont
  {Rice}},\ }\bibfield  {title} {\bibinfo {title} {{Effective Hamiltonian for
  the superconducting Cu oxides}},\ }\href
  {https://doi.org/10.1103/PhysRevB.37.3759} {\bibfield  {journal} {\bibinfo
  {journal} {Phys. Rev. B}\ }\textbf {\bibinfo {volume} {37}},\ \bibinfo
  {pages} {3759} (\bibinfo {year} {1988})}\BibitemShut {NoStop}%
\bibitem [{\citenamefont {Su}(1993)}]{tJSu}%
  \BibitemOpen
  \bibfield  {author} {\bibinfo {author} {\bibfnamefont {G.}~\bibnamefont
  {Su}},\ }\bibfield  {title} {\bibinfo {title} {{Remarks on the
  Brillouin-Wigner perturbation theory}},\ }\href
  {https://doi.org/10.1088/0305-4470/26/4/003} {\bibfield  {journal} {\bibinfo
  {journal} {Journal of Physics A: Mathematical and General}\ }\textbf
  {\bibinfo {volume} {26}},\ \bibinfo {pages} {L139} (\bibinfo {year}
  {1993})}\BibitemShut {NoStop}%
\bibitem [{\citenamefont {Zheng}\ \emph {et~al.}(2017)\citenamefont {Zheng},
  \citenamefont {Chung}, \citenamefont {Corboz}, \citenamefont {Ehlers},
  \citenamefont {Qin}, \citenamefont {Noack}, \citenamefont {Shi},
  \citenamefont {White}, \citenamefont {Zhang},\ and\ \citenamefont
  {Chan}}]{Zheng2017stripe}%
  \BibitemOpen
  \bibfield  {author} {\bibinfo {author} {\bibfnamefont {B.-X.}\ \bibnamefont
  {Zheng}}, \bibinfo {author} {\bibfnamefont {C.-M.}\ \bibnamefont {Chung}},
  \bibinfo {author} {\bibfnamefont {P.}~\bibnamefont {Corboz}}, \bibinfo
  {author} {\bibfnamefont {G.}~\bibnamefont {Ehlers}}, \bibinfo {author}
  {\bibfnamefont {M.-P.}\ \bibnamefont {Qin}}, \bibinfo {author} {\bibfnamefont
  {R.~M.}\ \bibnamefont {Noack}}, \bibinfo {author} {\bibfnamefont
  {H.}~\bibnamefont {Shi}}, \bibinfo {author} {\bibfnamefont {S.~R.}\
  \bibnamefont {White}}, \bibinfo {author} {\bibfnamefont {S.}~\bibnamefont
  {Zhang}},\ and\ \bibinfo {author} {\bibfnamefont {G.~K.-L.}\ \bibnamefont
  {Chan}},\ }\bibfield  {title} {\bibinfo {title} {{Stripe order in the
  underdoped region of the two-dimensional Hubbard model}},\ }\href
  {https://doi.org/10.1126/science.aam7127} {\bibfield  {journal} {\bibinfo
  {journal} {Science}\ }\textbf {\bibinfo {volume} {358}},\ \bibinfo {pages}
  {1155} (\bibinfo {year} {2017})}\BibitemShut {NoStop}%
\bibitem [{\citenamefont {Qin}\ \emph {et~al.}(2020)\citenamefont {Qin},
  \citenamefont {Chung}, \citenamefont {Shi}, \citenamefont {Vitali},
  \citenamefont {Hubig}, \citenamefont {Schollw\"ock}, \citenamefont {White},\
  and\ \citenamefont {Zhang}}]{AbsenceSCHubbard}%
  \BibitemOpen
  \bibfield  {author} {\bibinfo {author} {\bibfnamefont {M.}~\bibnamefont
  {Qin}}, \bibinfo {author} {\bibfnamefont {C.-M.}\ \bibnamefont {Chung}},
  \bibinfo {author} {\bibfnamefont {H.}~\bibnamefont {Shi}}, \bibinfo {author}
  {\bibfnamefont {E.}~\bibnamefont {Vitali}}, \bibinfo {author} {\bibfnamefont
  {C.}~\bibnamefont {Hubig}}, \bibinfo {author} {\bibfnamefont
  {U.}~\bibnamefont {Schollw\"ock}}, \bibinfo {author} {\bibfnamefont {S.~R.}\
  \bibnamefont {White}},\ and\ \bibinfo {author} {\bibfnamefont
  {S.}~\bibnamefont {Zhang}} (\bibinfo {collaboration} {Simons Collaboration on
  the Many-Electron Problem}),\ }\bibfield  {title} {\bibinfo {title} {{Absence
  of Superconductivity in the Pure Two-Dimensional Hubbard Model}},\ }\href
  {https://doi.org/10.1103/PhysRevX.10.031016} {\bibfield  {journal} {\bibinfo
  {journal} {Phys. Rev. X}\ }\textbf {\bibinfo {volume} {10}},\ \bibinfo
  {pages} {031016} (\bibinfo {year} {2020})}\BibitemShut {NoStop}%
\bibitem [{\citenamefont {Qin}\ \emph {et~al.}(2022)\citenamefont {Qin},
  \citenamefont {Sch\"{a}fer}, \citenamefont {Andergassen}, \citenamefont
  {Corboz},\ and\ \citenamefont {Gull}}]{MingpuQinHubbard2022}%
  \BibitemOpen
  \bibfield  {author} {\bibinfo {author} {\bibfnamefont {M.}~\bibnamefont
  {Qin}}, \bibinfo {author} {\bibfnamefont {T.}~\bibnamefont {Sch\"{a}fer}},
  \bibinfo {author} {\bibfnamefont {S.}~\bibnamefont {Andergassen}}, \bibinfo
  {author} {\bibfnamefont {P.}~\bibnamefont {Corboz}},\ and\ \bibinfo {author}
  {\bibfnamefont {E.}~\bibnamefont {Gull}},\ }\bibfield  {title} {\bibinfo
  {title} {{The Hubbard Model: A Computational Perspective}},\ }\href
  {https://doi.org/10.1146/annurev-conmatphys-090921-033948} {\bibfield
  {journal} {\bibinfo  {journal} {Annual Review of Condensed Matter Physics}\
  }\textbf {\bibinfo {volume} {13}},\ \bibinfo {pages} {275} (\bibinfo {year}
  {2022})}\BibitemShut {NoStop}%
\bibitem [{\citenamefont {Gong}\ \emph {et~al.}(2021)\citenamefont {Gong},
  \citenamefont {Zhu},\ and\ \citenamefont {Sheng}}]{Gong2021}%
  \BibitemOpen
  \bibfield  {author} {\bibinfo {author} {\bibfnamefont {S.}~\bibnamefont
  {Gong}}, \bibinfo {author} {\bibfnamefont {W.}~\bibnamefont {Zhu}},\ and\
  \bibinfo {author} {\bibfnamefont {D.~N.}\ \bibnamefont {Sheng}},\ }\bibfield
  {title} {\bibinfo {title} {{Robust $d$-Wave Superconductivity in the
  Square-Lattice $t\text{\ensuremath{-}}J$ Model}},\ }\href
  {https://doi.org/10.1103/PhysRevLett.127.097003} {\bibfield  {journal}
  {\bibinfo  {journal} {Phys. Rev. Lett.}\ }\textbf {\bibinfo {volume} {127}},\
  \bibinfo {pages} {097003} (\bibinfo {year} {2021})}\BibitemShut {NoStop}%
\bibitem [{\citenamefont {Jiang}\ \emph {et~al.}(2023)\citenamefont {Jiang},
  \citenamefont {Kivelson},\ and\ \citenamefont {Lee}}]{JiangHongchen8legtJ}%
  \BibitemOpen
  \bibfield  {author} {\bibinfo {author} {\bibfnamefont {H.-C.}\ \bibnamefont
  {Jiang}}, \bibinfo {author} {\bibfnamefont {S.~A.}\ \bibnamefont
  {Kivelson}},\ and\ \bibinfo {author} {\bibfnamefont {D.-H.}\ \bibnamefont
  {Lee}},\ }\bibfield  {title} {\bibinfo {title} {{Superconducting valence bond
  fluid in lightly doped eight-leg $t\text{\ensuremath{-}}J$ cylinders}},\
  }\href {https://doi.org/10.1103/PhysRevB.108.054505} {\bibfield  {journal}
  {\bibinfo  {journal} {Phys. Rev. B}\ }\textbf {\bibinfo {volume} {108}},\
  \bibinfo {pages} {054505} (\bibinfo {year} {2023})}\BibitemShut {NoStop}%
\bibitem [{\citenamefont {Jiang}\ and\ \citenamefont
  {Devereaux}(2019)}]{JiangHongchen2019}%
  \BibitemOpen
  \bibfield  {author} {\bibinfo {author} {\bibfnamefont {H.-C.}\ \bibnamefont
  {Jiang}}\ and\ \bibinfo {author} {\bibfnamefont {T.~P.}\ \bibnamefont
  {Devereaux}},\ }\bibfield  {title} {\bibinfo {title} {{Superconductivity in
  the doped Hubbard model and its interplay with next-nearest hopping
  ${t}^{\ensuremath{'}}$}},\ }\href {https://doi.org/10.1126/science.aal5304}
  {\bibfield  {journal} {\bibinfo  {journal} {Science}\ }\textbf {\bibinfo
  {volume} {365}},\ \bibinfo {pages} {1424} (\bibinfo {year}
  {2019})}\BibitemShut {NoStop}%
\bibitem [{\citenamefont {Jiang}\ \emph {et~al.}(2018)\citenamefont {Jiang},
  \citenamefont {Weng},\ and\ \citenamefont
  {Kivelson}}]{JiangHongchen2018-4leg}%
  \BibitemOpen
  \bibfield  {author} {\bibinfo {author} {\bibfnamefont {H.-C.}\ \bibnamefont
  {Jiang}}, \bibinfo {author} {\bibfnamefont {Z.-Y.}\ \bibnamefont {Weng}},\
  and\ \bibinfo {author} {\bibfnamefont {S.~A.}\ \bibnamefont {Kivelson}},\
  }\bibfield  {title} {\bibinfo {title} {{Superconductivity in the doped
  $\mathit{t}\ensuremath{-}\mathit{J}$ model: Results for four-leg
  cylinders}},\ }\href {https://doi.org/10.1103/PhysRevB.98.140505} {\bibfield
  {journal} {\bibinfo  {journal} {Phys. Rev. B}\ }\textbf {\bibinfo {volume}
  {98}},\ \bibinfo {pages} {140505} (\bibinfo {year} {2018})}\BibitemShut
  {NoStop}%
\bibitem [{\citenamefont {Jiang}\ \emph
  {et~al.}(2020{\natexlab{a}})\citenamefont {Jiang}, \citenamefont {Zaanen},
  \citenamefont {Devereaux},\ and\ \citenamefont {Jiang}}]{LE1LE2}%
  \BibitemOpen
  \bibfield  {author} {\bibinfo {author} {\bibfnamefont {Y.-F.}\ \bibnamefont
  {Jiang}}, \bibinfo {author} {\bibfnamefont {J.}~\bibnamefont {Zaanen}},
  \bibinfo {author} {\bibfnamefont {T.~P.}\ \bibnamefont {Devereaux}},\ and\
  \bibinfo {author} {\bibfnamefont {H.-C.}\ \bibnamefont {Jiang}},\ }\bibfield
  {title} {\bibinfo {title} {{Ground state phase diagram of the doped Hubbard
  model on the four-leg cylinder}},\ }\href
  {https://doi.org/10.1103/PhysRevResearch.2.033073} {\bibfield  {journal}
  {\bibinfo  {journal} {Phys. Rev. Res.}\ }\textbf {\bibinfo {volume} {2}},\
  \bibinfo {pages} {033073} (\bibinfo {year} {2020}{\natexlab{a}})}\BibitemShut
  {NoStop}%
\bibitem [{\citenamefont {Jiang}\ \emph {et~al.}(2021)\citenamefont {Jiang},
  \citenamefont {Scalapino},\ and\ \citenamefont {White}}]{White2021}%
  \BibitemOpen
  \bibfield  {author} {\bibinfo {author} {\bibfnamefont {S.}~\bibnamefont
  {Jiang}}, \bibinfo {author} {\bibfnamefont {D.~J.}\ \bibnamefont
  {Scalapino}},\ and\ \bibinfo {author} {\bibfnamefont {S.~R.}\ \bibnamefont
  {White}},\ }\bibfield  {title} {\bibinfo {title} {{Ground-state phase diagram
  of the $t\ensuremath{-}{t}^{\ensuremath{'}}\ensuremath{-}J$ model}},\ }\href
  {https://doi.org/10.1073/pnas.2109978118} {\bibfield  {journal} {\bibinfo
  {journal} {Proceedings of the National Academy of Sciences}\ }\textbf
  {\bibinfo {volume} {118}},\ \bibinfo {pages} {e2109978118} (\bibinfo {year}
  {2021})}\BibitemShut {NoStop}%
\bibitem [{\citenamefont {Qu}\ \emph {et~al.}(2024)\citenamefont {Qu},
  \citenamefont {Li}, \citenamefont {Gong}, \citenamefont {Qi}, \citenamefont
  {Li},\ and\ \citenamefont {Su}}]{David2022}%
  \BibitemOpen
  \bibfield  {author} {\bibinfo {author} {\bibfnamefont {D.-W.}\ \bibnamefont
  {Qu}}, \bibinfo {author} {\bibfnamefont {Q.}~\bibnamefont {Li}}, \bibinfo
  {author} {\bibfnamefont {S.-S.}\ \bibnamefont {Gong}}, \bibinfo {author}
  {\bibfnamefont {Y.}~\bibnamefont {Qi}}, \bibinfo {author} {\bibfnamefont
  {W.}~\bibnamefont {Li}},\ and\ \bibinfo {author} {\bibfnamefont
  {G.}~\bibnamefont {Su}},\ }\bibfield  {title} {\bibinfo {title} {{Phase
  Diagram, $d$-Wave Superconductivity, and Pseudogap of the
  $t\ensuremath{-}{t}^{\ensuremath{'}}\ensuremath{-}J$ Model at Finite
  Temperature}},\ }\href {https://doi.org/10.1103/PhysRevLett.133.256003}
  {\bibfield  {journal} {\bibinfo  {journal} {Phys. Rev. Lett.}\ }\textbf
  {\bibinfo {volume} {133}},\ \bibinfo {pages} {256003} (\bibinfo {year}
  {2024})}\BibitemShut {NoStop}%
\bibitem [{\citenamefont {Chung}\ \emph {et~al.}(2020)\citenamefont {Chung},
  \citenamefont {Qin}, \citenamefont {Zhang}, \citenamefont {Schollw\"ock},\
  and\ \citenamefont {White}}]{Plaq_d_wave}%
  \BibitemOpen
  \bibfield  {author} {\bibinfo {author} {\bibfnamefont {C.-M.}\ \bibnamefont
  {Chung}}, \bibinfo {author} {\bibfnamefont {M.}~\bibnamefont {Qin}}, \bibinfo
  {author} {\bibfnamefont {S.}~\bibnamefont {Zhang}}, \bibinfo {author}
  {\bibfnamefont {U.}~\bibnamefont {Schollw\"ock}},\ and\ \bibinfo {author}
  {\bibfnamefont {S.~R.}\ \bibnamefont {White}} (\bibinfo {collaboration} {The
  Simons Collaboration on the Many-Electron Problem}),\ }\bibfield  {title}
  {\bibinfo {title} {{Plaquette versus ordinary $d$-wave pairing in the
  ${t}^{\ensuremath{'}}$-Hubbard model on a width-4 cylinder}},\ }\href
  {https://doi.org/10.1103/PhysRevB.102.041106} {\bibfield  {journal} {\bibinfo
   {journal} {Phys. Rev. B}\ }\textbf {\bibinfo {volume} {102}},\ \bibinfo
  {pages} {041106} (\bibinfo {year} {2020})}\BibitemShut {NoStop}%
\bibitem [{\citenamefont {Jiang}\ \emph
  {et~al.}(2020{\natexlab{b}})\citenamefont {Jiang}, \citenamefont {Chen},\
  and\ \citenamefont {Weng}}]{LELPhaseStringtJ}%
  \BibitemOpen
  \bibfield  {author} {\bibinfo {author} {\bibfnamefont {H.-C.}\ \bibnamefont
  {Jiang}}, \bibinfo {author} {\bibfnamefont {S.}~\bibnamefont {Chen}},\ and\
  \bibinfo {author} {\bibfnamefont {Z.-Y.}\ \bibnamefont {Weng}},\ }\bibfield
  {title} {\bibinfo {title} {{Critical role of the sign structure in the doped
  Mott insulator: Luther-Emery versus Fermi-liquid-like state in
  quasi-one-dimensional ladders}},\ }\href
  {https://doi.org/10.1103/PhysRevB.102.104512} {\bibfield  {journal} {\bibinfo
   {journal} {Phys. Rev. B}\ }\textbf {\bibinfo {volume} {102}},\ \bibinfo
  {pages} {104512} (\bibinfo {year} {2020}{\natexlab{b}})}\BibitemShut
  {NoStop}%
\bibitem [{\citenamefont {Hamidian}\ \emph {et~al.}(2016)\citenamefont
  {Hamidian}, \citenamefont {Edkins}, \citenamefont {Joo}, \citenamefont
  {Kostin}, \citenamefont {Eisaki}, \citenamefont {Uchida}, \citenamefont
  {Lawler}, \citenamefont {Kim}, \citenamefont {Mackenzie}, \citenamefont
  {Fujita}, \citenamefont {Lee},\ and\ \citenamefont {Davis}}]{PDWBSCCO}%
  \BibitemOpen
  \bibfield  {author} {\bibinfo {author} {\bibfnamefont {M.~H.}\ \bibnamefont
  {Hamidian}}, \bibinfo {author} {\bibfnamefont {S.~D.}\ \bibnamefont
  {Edkins}}, \bibinfo {author} {\bibfnamefont {S.~H.}\ \bibnamefont {Joo}},
  \bibinfo {author} {\bibfnamefont {A.}~\bibnamefont {Kostin}}, \bibinfo
  {author} {\bibfnamefont {H.}~\bibnamefont {Eisaki}}, \bibinfo {author}
  {\bibfnamefont {S.}~\bibnamefont {Uchida}}, \bibinfo {author} {\bibfnamefont
  {M.~J.}\ \bibnamefont {Lawler}}, \bibinfo {author} {\bibfnamefont {E.-A.}\
  \bibnamefont {Kim}}, \bibinfo {author} {\bibfnamefont {A.~P.}\ \bibnamefont
  {Mackenzie}}, \bibinfo {author} {\bibfnamefont {K.}~\bibnamefont {Fujita}},
  \bibinfo {author} {\bibfnamefont {J.}~\bibnamefont {Lee}},\ and\ \bibinfo
  {author} {\bibfnamefont {J.~C.~S.}\ \bibnamefont {Davis}},\ }\bibfield
  {title} {\bibinfo {title} {{Detection of a Cooper-pair density wave in
  Bi$_2$Sr$_2$CaCu$_2$O$_{8+x}$}},\ }\href
  {https://doi.org/10.1038/nature17411} {\bibfield  {journal} {\bibinfo
  {journal} {Nature}\ }\textbf {\bibinfo {volume} {532}},\ \bibinfo {pages}
  {343} (\bibinfo {year} {2016})}\BibitemShut {NoStop}%
\bibitem [{\citenamefont {Du}\ \emph {et~al.}(2020)\citenamefont {Du},
  \citenamefont {Li}, \citenamefont {Joo}, \citenamefont {Donoway},
  \citenamefont {Lee}, \citenamefont {Davis}, \citenamefont {Gu}, \citenamefont
  {Johnson},\ and\ \citenamefont {Fujita}}]{PDWcuprateFujita}%
  \BibitemOpen
  \bibfield  {author} {\bibinfo {author} {\bibfnamefont {Z.}~\bibnamefont
  {Du}}, \bibinfo {author} {\bibfnamefont {H.}~\bibnamefont {Li}}, \bibinfo
  {author} {\bibfnamefont {S.~H.}\ \bibnamefont {Joo}}, \bibinfo {author}
  {\bibfnamefont {E.~P.}\ \bibnamefont {Donoway}}, \bibinfo {author}
  {\bibfnamefont {J.}~\bibnamefont {Lee}}, \bibinfo {author} {\bibfnamefont
  {J.~C.~S.}\ \bibnamefont {Davis}}, \bibinfo {author} {\bibfnamefont
  {G.}~\bibnamefont {Gu}}, \bibinfo {author} {\bibfnamefont {P.~D.}\
  \bibnamefont {Johnson}},\ and\ \bibinfo {author} {\bibfnamefont
  {K.}~\bibnamefont {Fujita}},\ }\bibfield  {title} {\bibinfo {title} {{Imaging
  the energy gap modulations of the cuprate pair-density-wave state}},\ }\href
  {https://doi.org/10.1038/s41586-020-2143-x} {\bibfield  {journal} {\bibinfo
  {journal} {Nature}\ }\textbf {\bibinfo {volume} {580}},\ \bibinfo {pages}
  {65} (\bibinfo {year} {2020})}\BibitemShut {NoStop}%
\bibitem [{\citenamefont {Edkins}\ \emph {et~al.}(2019)\citenamefont {Edkins},
  \citenamefont {Kostin}, \citenamefont {Fujita}, \citenamefont {Mackenzie},
  \citenamefont {Eisaki}, \citenamefont {Uchida}, \citenamefont {Sachdev},
  \citenamefont {Lawler}, \citenamefont {Kim}, \citenamefont {Davis},\ and\
  \citenamefont {Hamidian}}]{PDWcuprateHamidian}%
  \BibitemOpen
  \bibfield  {author} {\bibinfo {author} {\bibfnamefont {S.~D.}\ \bibnamefont
  {Edkins}}, \bibinfo {author} {\bibfnamefont {A.}~\bibnamefont {Kostin}},
  \bibinfo {author} {\bibfnamefont {K.}~\bibnamefont {Fujita}}, \bibinfo
  {author} {\bibfnamefont {A.~P.}\ \bibnamefont {Mackenzie}}, \bibinfo {author}
  {\bibfnamefont {H.}~\bibnamefont {Eisaki}}, \bibinfo {author} {\bibfnamefont
  {S.}~\bibnamefont {Uchida}}, \bibinfo {author} {\bibfnamefont
  {S.}~\bibnamefont {Sachdev}}, \bibinfo {author} {\bibfnamefont {M.~J.}\
  \bibnamefont {Lawler}}, \bibinfo {author} {\bibfnamefont {E.-A.}\
  \bibnamefont {Kim}}, \bibinfo {author} {\bibfnamefont {J.~C.~S.}\
  \bibnamefont {Davis}},\ and\ \bibinfo {author} {\bibfnamefont {M.~H.}\
  \bibnamefont {Hamidian}},\ }\bibfield  {title} {\bibinfo {title} {{Magnetic
  field–induced pair density wave state in the cuprate vortex halo}},\ }\href
  {https://doi.org/10.1126/science.aat1773} {\bibfield  {journal} {\bibinfo
  {journal} {Science}\ }\textbf {\bibinfo {volume} {364}},\ \bibinfo {pages}
  {976} (\bibinfo {year} {2019})}\BibitemShut {NoStop}%
\bibitem [{\citenamefont {Ruan}\ \emph {et~al.}(2018)\citenamefont {Ruan},
  \citenamefont {Li}, \citenamefont {Hu}, \citenamefont {Hao}, \citenamefont
  {Li}, \citenamefont {Cai}, \citenamefont {Zhou}, \citenamefont {Lee},\ and\
  \citenamefont {Wang}}]{PDWcuprateWang}%
  \BibitemOpen
  \bibfield  {author} {\bibinfo {author} {\bibfnamefont {W.}~\bibnamefont
  {Ruan}}, \bibinfo {author} {\bibfnamefont {X.}~\bibnamefont {Li}}, \bibinfo
  {author} {\bibfnamefont {C.}~\bibnamefont {Hu}}, \bibinfo {author}
  {\bibfnamefont {Z.}~\bibnamefont {Hao}}, \bibinfo {author} {\bibfnamefont
  {H.}~\bibnamefont {Li}}, \bibinfo {author} {\bibfnamefont {P.}~\bibnamefont
  {Cai}}, \bibinfo {author} {\bibfnamefont {X.}~\bibnamefont {Zhou}}, \bibinfo
  {author} {\bibfnamefont {D.-H.}\ \bibnamefont {Lee}},\ and\ \bibinfo {author}
  {\bibfnamefont {Y.}~\bibnamefont {Wang}},\ }\bibfield  {title} {\bibinfo
  {title} {Visualization of the periodic modulation of cooper pairing in a
  cuprate superconductor},\ }\href {https://doi.org/10.1038/s41567-018-0276-8}
  {\bibfield  {journal} {\bibinfo  {journal} {Nature Physics}\ }\textbf
  {\bibinfo {volume} {14}},\ \bibinfo {pages} {1178} (\bibinfo {year}
  {2018})}\BibitemShut {NoStop}%
\bibitem [{\citenamefont {Agterberg}\ \emph {et~al.}(2020)\citenamefont
  {Agterberg}, \citenamefont {Davis}, \citenamefont {Edkins}, \citenamefont
  {Fradkin}, \citenamefont {Van~Harlingen}, \citenamefont {Kivelson},
  \citenamefont {Lee}, \citenamefont {Radzihovsky}, \citenamefont {Tranquada},\
  and\ \citenamefont {Wang}}]{PDWAnnualRev}%
  \BibitemOpen
  \bibfield  {author} {\bibinfo {author} {\bibfnamefont {D.~F.}\ \bibnamefont
  {Agterberg}}, \bibinfo {author} {\bibfnamefont {J.~S.}\ \bibnamefont
  {Davis}}, \bibinfo {author} {\bibfnamefont {S.~D.}\ \bibnamefont {Edkins}},
  \bibinfo {author} {\bibfnamefont {E.}~\bibnamefont {Fradkin}}, \bibinfo
  {author} {\bibfnamefont {D.~J.}\ \bibnamefont {Van~Harlingen}}, \bibinfo
  {author} {\bibfnamefont {S.~A.}\ \bibnamefont {Kivelson}}, \bibinfo {author}
  {\bibfnamefont {P.~A.}\ \bibnamefont {Lee}}, \bibinfo {author} {\bibfnamefont
  {L.}~\bibnamefont {Radzihovsky}}, \bibinfo {author} {\bibfnamefont {J.~M.}\
  \bibnamefont {Tranquada}},\ and\ \bibinfo {author} {\bibfnamefont
  {Y.}~\bibnamefont {Wang}},\ }\bibfield  {title} {\bibinfo {title} {{The
  Physics of Pair-Density Waves: Cuprate Superconductors and Beyond}},\ }\href
  {https://doi.org/10.1146/annurev-conmatphys-031119-050711} {\bibfield
  {journal} {\bibinfo  {journal} {Annual Review of Condensed Matter Physics}\
  }\textbf {\bibinfo {volume} {11}},\ \bibinfo {pages} {231} (\bibinfo {year}
  {2020})}\BibitemShut {NoStop}%
\bibitem [{\citenamefont {Jiang}\ and\ \citenamefont
  {Devereaux}(2023)}]{JiangHongchen3band}%
  \BibitemOpen
  \bibfield  {author} {\bibinfo {author} {\bibfnamefont {H.-C.}\ \bibnamefont
  {Jiang}}\ and\ \bibinfo {author} {\bibfnamefont {T.~P.}\ \bibnamefont
  {Devereaux}},\ }\bibfield  {title} {\bibinfo {title} {Pair density wave and
  superconductivity in a kinetically frustrated doped emery model on a square
  lattice},\ }\bibfield  {journal} {\bibinfo  {journal} {Frontiers in
  Electronic Materials}\ }\textbf {\bibinfo {volume} {3}},\ \href
  {https://doi.org/10.3389/femat.2023.1323404} {10.3389/femat.2023.1323404}
  (\bibinfo {year} {2023})\BibitemShut {NoStop}%
\bibitem [{\citenamefont {Zhang}\ \emph {et~al.}(2023)\citenamefont {Zhang},
  \citenamefont {Sun},\ and\ \citenamefont {Weng}}]{ZhangPDW2023}%
  \BibitemOpen
  \bibfield  {author} {\bibinfo {author} {\bibfnamefont {H.-K.}\ \bibnamefont
  {Zhang}}, \bibinfo {author} {\bibfnamefont {R.-Y.}\ \bibnamefont {Sun}},\
  and\ \bibinfo {author} {\bibfnamefont {Z.-Y.}\ \bibnamefont {Weng}},\
  }\bibfield  {title} {\bibinfo {title} {{Pair density wave characterized by a
  hidden string order parameter}},\ }\href
  {https://doi.org/10.1103/PhysRevB.108.115136} {\bibfield  {journal} {\bibinfo
   {journal} {Phys. Rev. B}\ }\textbf {\bibinfo {volume} {108}},\ \bibinfo
  {pages} {115136} (\bibinfo {year} {2023})}\BibitemShut {NoStop}%
\bibitem [{\citenamefont {Zheng}\ \emph {et~al.}(2025)\citenamefont {Zheng},
  \citenamefont {Yue}, \citenamefont {Zhang},\ and\ \citenamefont
  {Gu}}]{PDWpeps}%
  \BibitemOpen
  \bibfield  {author} {\bibinfo {author} {\bibfnamefont {W.}~\bibnamefont
  {Zheng}}, \bibinfo {author} {\bibfnamefont {Z.-Y.}\ \bibnamefont {Yue}},
  \bibinfo {author} {\bibfnamefont {J.-H.}\ \bibnamefont {Zhang}},\ and\
  \bibinfo {author} {\bibfnamefont {Z.-C.}\ \bibnamefont {Gu}},\ }\bibfield
  {title} {\bibinfo {title} {{Competing pair density wave orders in the square
  lattice $t$-$J$ model}},\ }\href {https://doi.org/10.1038/s42005-025-02346-0}
  {\bibfield  {journal} {\bibinfo  {journal} {Communications Physics}\ }\textbf
  {\bibinfo {volume} {8}},\ \bibinfo {pages} {456} (\bibinfo {year}
  {2025})}\BibitemShut {NoStop}%
\bibitem [{\citenamefont {Fulde}\ and\ \citenamefont {Ferrell}(1964)}]{FF}%
  \BibitemOpen
  \bibfield  {author} {\bibinfo {author} {\bibfnamefont {P.}~\bibnamefont
  {Fulde}}\ and\ \bibinfo {author} {\bibfnamefont {R.~A.}\ \bibnamefont
  {Ferrell}},\ }\bibfield  {title} {\bibinfo {title} {{Superconductivity in a
  Strong Spin-Exchange Field}},\ }\href
  {https://doi.org/10.1103/PhysRev.135.A550} {\bibfield  {journal} {\bibinfo
  {journal} {Phys. Rev.}\ }\textbf {\bibinfo {volume} {135}},\ \bibinfo {pages}
  {A550} (\bibinfo {year} {1964})}\BibitemShut {NoStop}%
\bibitem [{\citenamefont {Larkin}\ and\ \citenamefont
  {Ovchinnikov}(1965)}]{LO}%
  \BibitemOpen
  \bibfield  {author} {\bibinfo {author} {\bibfnamefont {A.}~\bibnamefont
  {Larkin}}\ and\ \bibinfo {author} {\bibfnamefont {Y.~N.}\ \bibnamefont
  {Ovchinnikov}},\ }\bibfield  {title} {\bibinfo {title} {{Nonuniform state of
  superconductors}},\ }\href@noop {} {\bibfield  {journal} {\bibinfo  {journal}
  {Soviet Physics-JETP}\ }\textbf {\bibinfo {volume} {20}},\ \bibinfo {pages}
  {762} (\bibinfo {year} {1965})}\BibitemShut {NoStop}%
\bibitem [{\citenamefont {Tanamoto}\ \emph {et~al.}(1991)\citenamefont
  {Tanamoto}, \citenamefont {Kuboki},\ and\ \citenamefont
  {Fukuyama}}]{tJMagSlaveBoson}%
  \BibitemOpen
  \bibfield  {author} {\bibinfo {author} {\bibfnamefont {T.}~\bibnamefont
  {Tanamoto}}, \bibinfo {author} {\bibfnamefont {K.}~\bibnamefont {Kuboki}},\
  and\ \bibinfo {author} {\bibfnamefont {H.}~\bibnamefont {Fukuyama}},\
  }\bibfield  {title} {\bibinfo {title} {{Magnetic properties of $t$-$J$
  model}},\ }\href {https://doi.org/10.1143/JPSJ.60.3072} {\bibfield  {journal}
  {\bibinfo  {journal} {Journal of the Physical Society of Japan}\ }\textbf
  {\bibinfo {volume} {60}},\ \bibinfo {pages} {3072} (\bibinfo {year}
  {1991})}\BibitemShut {NoStop}%
\bibitem [{\citenamefont {Albuquerque}\ and\ \citenamefont
  {Martins}(2005)}]{tJMagED}%
  \BibitemOpen
  \bibfield  {author} {\bibinfo {author} {\bibfnamefont {A.~F.}\ \bibnamefont
  {Albuquerque}}\ and\ \bibinfo {author} {\bibfnamefont {G.~B.}\ \bibnamefont
  {Martins}},\ }\bibfield  {title} {\bibinfo {title} {{New results for the t-J
  model in ladders: changes in the spin liquid state with applied magnetic
  field; implications for the cuprates}},\ }\href
  {https://doi.org/10.1088/0953-8984/17/15/013} {\bibfield  {journal} {\bibinfo
   {journal} {Journal of Physics: Condensed Matter}\ }\textbf {\bibinfo
  {volume} {17}},\ \bibinfo {pages} {2419} (\bibinfo {year}
  {2005})}\BibitemShut {NoStop}%
\bibitem [{\citenamefont {Roux}\ \emph {et~al.}(2007)\citenamefont {Roux},
  \citenamefont {Orignac}, \citenamefont {White},\ and\ \citenamefont
  {Poilblanc}}]{tJladderflux}%
  \BibitemOpen
  \bibfield  {author} {\bibinfo {author} {\bibfnamefont {G.}~\bibnamefont
  {Roux}}, \bibinfo {author} {\bibfnamefont {E.}~\bibnamefont {Orignac}},
  \bibinfo {author} {\bibfnamefont {S.~R.}\ \bibnamefont {White}},\ and\
  \bibinfo {author} {\bibfnamefont {D.}~\bibnamefont {Poilblanc}},\ }\bibfield
  {title} {\bibinfo {title} {{Diamagnetism of doped two-leg ladders and probing
  the nature of their commensurate phases}},\ }\href
  {https://doi.org/10.1103/PhysRevB.76.195105} {\bibfield  {journal} {\bibinfo
  {journal} {Phys. Rev. B}\ }\textbf {\bibinfo {volume} {76}},\ \bibinfo
  {pages} {195105} (\bibinfo {year} {2007})}\BibitemShut {NoStop}%
\bibitem [{\citenamefont {Baldelli}\ \emph {et~al.}(2025)\citenamefont
  {Baldelli}, \citenamefont {Karlsson}, \citenamefont {Kloss}, \citenamefont
  {Fishman},\ and\ \citenamefont {Wietek}}]{Wietekflux}%
  \BibitemOpen
  \bibfield  {author} {\bibinfo {author} {\bibfnamefont {N.}~\bibnamefont
  {Baldelli}}, \bibinfo {author} {\bibfnamefont {H.}~\bibnamefont {Karlsson}},
  \bibinfo {author} {\bibfnamefont {B.}~\bibnamefont {Kloss}}, \bibinfo
  {author} {\bibfnamefont {M.}~\bibnamefont {Fishman}},\ and\ \bibinfo {author}
  {\bibfnamefont {A.}~\bibnamefont {Wietek}},\ }\bibfield  {title} {\bibinfo
  {title} {{Fragmented superconductivity in the Hubbard model as solitons in
  Ginzburg--Landau theory}},\ }\href
  {https://doi.org/10.1038/s41535-024-00718-3} {\bibfield  {journal} {\bibinfo
  {journal} {npj Quantum Materials}\ }\textbf {\bibinfo {volume} {10}},\
  \bibinfo {pages} {22} (\bibinfo {year} {2025})}\BibitemShut {NoStop}%
\bibitem [{\citenamefont {Pruschke}\ and\ \citenamefont
  {Shiba}(1992{\natexlab{a}})}]{1DtJMag}%
  \BibitemOpen
  \bibfield  {author} {\bibinfo {author} {\bibfnamefont {T.}~\bibnamefont
  {Pruschke}}\ and\ \bibinfo {author} {\bibfnamefont {H.}~\bibnamefont
  {Shiba}},\ }\bibfield  {title} {\bibinfo {title} {{Superconducting
  correlations in the one-dimensional t-J model}},\ }\href
  {https://doi.org/10.1103/PhysRevB.46.356} {\bibfield  {journal} {\bibinfo
  {journal} {Phys. Rev. B}\ }\textbf {\bibinfo {volume} {46}},\ \bibinfo
  {pages} {356} (\bibinfo {year} {1992}{\natexlab{a}})}\BibitemShut {NoStop}%
\bibitem [{\citenamefont {Roux}\ \emph {et~al.}(2006)\citenamefont {Roux},
  \citenamefont {White}, \citenamefont {Capponi},\ and\ \citenamefont
  {Poilblanc}}]{White2006}%
  \BibitemOpen
  \bibfield  {author} {\bibinfo {author} {\bibfnamefont {G.}~\bibnamefont
  {Roux}}, \bibinfo {author} {\bibfnamefont {S.~R.}\ \bibnamefont {White}},
  \bibinfo {author} {\bibfnamefont {S.}~\bibnamefont {Capponi}},\ and\ \bibinfo
  {author} {\bibfnamefont {D.}~\bibnamefont {Poilblanc}},\ }\bibfield  {title}
  {\bibinfo {title} {{Zeeman Effect in Superconducting Two-Leg Ladders:
  Irrational Magnetization Plateaus and Exceeding the Pauli Limit}},\ }\href
  {https://doi.org/10.1103/PhysRevLett.97.087207} {\bibfield  {journal}
  {\bibinfo  {journal} {Phys. Rev. Lett.}\ }\textbf {\bibinfo {volume} {97}},\
  \bibinfo {pages} {087207} (\bibinfo {year} {2006})}\BibitemShut {NoStop}%
\bibitem [{\citenamefont {Proust}\ and\ \citenamefont
  {Taillefer}(2019)}]{CuprateMagAnnualRev}%
  \BibitemOpen
  \bibfield  {author} {\bibinfo {author} {\bibfnamefont {C.}~\bibnamefont
  {Proust}}\ and\ \bibinfo {author} {\bibfnamefont {L.}~\bibnamefont
  {Taillefer}},\ }\bibfield  {title} {\bibinfo {title} {{The Remarkable
  Underlying Ground States of Cuprate Superconductors}},\ }\href
  {https://doi.org/10.1146/annurev-conmatphys-031218-013210} {\bibfield
  {journal} {\bibinfo  {journal} {Annual Review of Condensed Matter Physics}\
  }\textbf {\bibinfo {volume} {10}},\ \bibinfo {pages} {409} (\bibinfo {year}
  {2019})}\BibitemShut {NoStop}%
\bibitem [{\citenamefont {Helfand}\ and\ \citenamefont
  {Werthamer}(1966)}]{OrbitalEff1}%
  \BibitemOpen
  \bibfield  {author} {\bibinfo {author} {\bibfnamefont {E.}~\bibnamefont
  {Helfand}}\ and\ \bibinfo {author} {\bibfnamefont {N.~R.}\ \bibnamefont
  {Werthamer}},\ }\bibfield  {title} {\bibinfo {title} {{Temperature and Purity
  Dependence of the Superconducting Critical Field, ${H}_{c2}$. II}},\ }\href
  {https://doi.org/10.1103/PhysRev.147.288} {\bibfield  {journal} {\bibinfo
  {journal} {Phys. Rev.}\ }\textbf {\bibinfo {volume} {147}},\ \bibinfo {pages}
  {288} (\bibinfo {year} {1966})}\BibitemShut {NoStop}%
\bibitem [{\citenamefont {Werthamer}\ \emph {et~al.}(1966)\citenamefont
  {Werthamer}, \citenamefont {Helfand},\ and\ \citenamefont
  {Hohenberg}}]{OrbitalEff2}%
  \BibitemOpen
  \bibfield  {author} {\bibinfo {author} {\bibfnamefont {N.~R.}\ \bibnamefont
  {Werthamer}}, \bibinfo {author} {\bibfnamefont {E.}~\bibnamefont {Helfand}},\
  and\ \bibinfo {author} {\bibfnamefont {P.~C.}\ \bibnamefont {Hohenberg}},\
  }\bibfield  {title} {\bibinfo {title} {{Temperature and Purity Dependence of
  the Superconducting Critical Field, ${H}_{c2}$. III. Electron Spin and
  Spin-Orbit Effects}},\ }\href {https://doi.org/10.1103/PhysRev.147.295}
  {\bibfield  {journal} {\bibinfo  {journal} {Phys. Rev.}\ }\textbf {\bibinfo
  {volume} {147}},\ \bibinfo {pages} {295} (\bibinfo {year}
  {1966})}\BibitemShut {NoStop}%
\bibitem [{\citenamefont {Sarma}(1963)}]{ZeemanEff1}%
  \BibitemOpen
  \bibfield  {author} {\bibinfo {author} {\bibfnamefont {G.}~\bibnamefont
  {Sarma}},\ }\bibfield  {title} {\bibinfo {title} {{On the influence of a
  uniform exchange field acting on the spins of the conduction electrons in a
  superconductor}},\ }\href
  {https://doi.org/https://doi.org/10.1016/0022-3697(63)90007-6} {\bibfield
  {journal} {\bibinfo  {journal} {Journal of Physics and Chemistry of Solids}\
  }\textbf {\bibinfo {volume} {24}},\ \bibinfo {pages} {1029} (\bibinfo {year}
  {1963})}\BibitemShut {NoStop}%
\bibitem [{\citenamefont {Maki}\ and\ \citenamefont
  {Tsuneto}(1964)}]{MakiTransitionOrder}%
  \BibitemOpen
  \bibfield  {author} {\bibinfo {author} {\bibfnamefont {K.}~\bibnamefont
  {Maki}}\ and\ \bibinfo {author} {\bibfnamefont {T.}~\bibnamefont {Tsuneto}},\
  }\bibfield  {title} {\bibinfo {title} {{Pauli paramagnetism and
  superconducting state}},\ }\href {https://doi.org/10.1143/PTP.31.945}
  {\bibfield  {journal} {\bibinfo  {journal} {Progress of Theoretical Physics}\
  }\textbf {\bibinfo {volume} {31}},\ \bibinfo {pages} {945} (\bibinfo {year}
  {1964})}\BibitemShut {NoStop}%
\bibitem [{\citenamefont {Clogston}(1962)}]{PauliLimit}%
  \BibitemOpen
  \bibfield  {author} {\bibinfo {author} {\bibfnamefont {A.~M.}\ \bibnamefont
  {Clogston}},\ }\bibfield  {title} {\bibinfo {title} {{Upper Limit for the
  Critical Field in Hard Superconductors}},\ }\href
  {https://doi.org/10.1103/PhysRevLett.9.266} {\bibfield  {journal} {\bibinfo
  {journal} {Phys. Rev. Lett.}\ }\textbf {\bibinfo {volume} {9}},\ \bibinfo
  {pages} {266} (\bibinfo {year} {1962})}\BibitemShut {NoStop}%
\bibitem [{\citenamefont {Chandrasekhar}(1962)}]{PauliLimitChandrasekhar}%
  \BibitemOpen
  \bibfield  {author} {\bibinfo {author} {\bibfnamefont {B.~S.}\ \bibnamefont
  {Chandrasekhar}},\ }\bibfield  {title} {\bibinfo {title} {A note on the
  maximum critical field of high-field superconductors},\ }\href
  {https://doi.org/10.1063/1.1777362} {\bibfield  {journal} {\bibinfo
  {journal} {Applied Physics Letters}\ }\textbf {\bibinfo {volume} {1}},\
  \bibinfo {pages} {7} (\bibinfo {year} {1962})}\BibitemShut {NoStop}%
\bibitem [{\citenamefont {Agosta}(2018)}]{cryst8070285}%
  \BibitemOpen
  \bibfield  {author} {\bibinfo {author} {\bibfnamefont {C.~C.}\ \bibnamefont
  {Agosta}},\ }\bibfield  {title} {\bibinfo {title} {{Inhomogeneous
  Superconductivity in Organic and Related Superconductors}},\ }\bibfield
  {journal} {\bibinfo  {journal} {Crystals}\ }\textbf {\bibinfo {volume} {8}},\
  \href {https://doi.org/10.3390/cryst8070285} {10.3390/cryst8070285} (\bibinfo
  {year} {2018})\BibitemShut {NoStop}%
\bibitem [{\citenamefont {Gruenberg}\ and\ \citenamefont
  {Gunther}(1966)}]{Makibound}%
  \BibitemOpen
  \bibfield  {author} {\bibinfo {author} {\bibfnamefont {L.~W.}\ \bibnamefont
  {Gruenberg}}\ and\ \bibinfo {author} {\bibfnamefont {L.}~\bibnamefont
  {Gunther}},\ }\bibfield  {title} {\bibinfo {title} {{Fulde-Ferrell Effect in
  Type-II Superconductors}},\ }\href
  {https://doi.org/10.1103/PhysRevLett.16.996} {\bibfield  {journal} {\bibinfo
  {journal} {Phys. Rev. Lett.}\ }\textbf {\bibinfo {volume} {16}},\ \bibinfo
  {pages} {996} (\bibinfo {year} {1966})}\BibitemShut {NoStop}%
\bibitem [{\citenamefont {Maki}(1966)}]{Maki1966}%
  \BibitemOpen
  \bibfield  {author} {\bibinfo {author} {\bibfnamefont {K.}~\bibnamefont
  {Maki}},\ }\bibfield  {title} {\bibinfo {title} {{Effect of Pauli
  Paramagnetism on Magnetic Properties of High-Field Superconductors}},\ }\href
  {https://doi.org/10.1103/PhysRev.148.362} {\bibfield  {journal} {\bibinfo
  {journal} {Phys. Rev.}\ }\textbf {\bibinfo {volume} {148}},\ \bibinfo {pages}
  {362} (\bibinfo {year} {1966})}\BibitemShut {NoStop}%
\bibitem [{\citenamefont {Mayaffre}\ \emph {et~al.}(2014)\citenamefont
  {Mayaffre}, \citenamefont {Kr{\"a}mer}, \citenamefont {Horvatić},
  \citenamefont {Berthier}, \citenamefont {Miyagawa}, \citenamefont {Kanoda},\
  and\ \citenamefont {Mitrović}}]{OrganicAndreev}%
  \BibitemOpen
  \bibfield  {author} {\bibinfo {author} {\bibfnamefont {H.}~\bibnamefont
  {Mayaffre}}, \bibinfo {author} {\bibfnamefont {S.}~\bibnamefont
  {Kr{\"a}mer}}, \bibinfo {author} {\bibfnamefont {M.}~\bibnamefont
  {Horvatić}}, \bibinfo {author} {\bibfnamefont {C.}~\bibnamefont {Berthier}},
  \bibinfo {author} {\bibfnamefont {K.}~\bibnamefont {Miyagawa}}, \bibinfo
  {author} {\bibfnamefont {K.}~\bibnamefont {Kanoda}},\ and\ \bibinfo {author}
  {\bibfnamefont {V.~F.}\ \bibnamefont {Mitrović}},\ }\bibfield  {title}
  {\bibinfo {title} {{Evidence of Andreev bound states as a hallmark of the
  FFLO phase in $\kappa$-(BEDT-TTF)$_2$Cu(NCS)$_2$}},\ }\href
  {https://doi.org/10.1038/nphys3121} {\bibfield  {journal} {\bibinfo
  {journal} {Nature Physics}\ }\textbf {\bibinfo {volume} {10}},\ \bibinfo
  {pages} {928–932} (\bibinfo {year} {2014})}\BibitemShut {NoStop}%
\bibitem [{\citenamefont {Lortz}\ \emph {et~al.}(2007)\citenamefont {Lortz},
  \citenamefont {Wang}, \citenamefont {Demuer}, \citenamefont {B\"ottger},
  \citenamefont {Bergk}, \citenamefont {Zwicknagl}, \citenamefont {Nakazawa},\
  and\ \citenamefont {Wosnitza}}]{OrganicCalori}%
  \BibitemOpen
  \bibfield  {author} {\bibinfo {author} {\bibfnamefont {R.}~\bibnamefont
  {Lortz}}, \bibinfo {author} {\bibfnamefont {Y.}~\bibnamefont {Wang}},
  \bibinfo {author} {\bibfnamefont {A.}~\bibnamefont {Demuer}}, \bibinfo
  {author} {\bibfnamefont {P.~H.~M.}\ \bibnamefont {B\"ottger}}, \bibinfo
  {author} {\bibfnamefont {B.}~\bibnamefont {Bergk}}, \bibinfo {author}
  {\bibfnamefont {G.}~\bibnamefont {Zwicknagl}}, \bibinfo {author}
  {\bibfnamefont {Y.}~\bibnamefont {Nakazawa}},\ and\ \bibinfo {author}
  {\bibfnamefont {J.}~\bibnamefont {Wosnitza}},\ }\bibfield  {title} {\bibinfo
  {title} {{Calorimetric Evidence for a Fulde-Ferrell-Larkin-Ovchinnikov
  Superconducting State in the Layered Organic Superconductor
  $\ensuremath{\kappa}\mathrm{\text{\ensuremath{-}}}(\mathrm{BEDT}\mathrm{\text{\ensuremath{-}}}\mathrm{TTF}{)}_{2}\mathrm{Cu}(\mathrm{NCS}{)}_{2}$}},\
  }\href {https://doi.org/10.1103/PhysRevLett.99.187002} {\bibfield  {journal}
  {\bibinfo  {journal} {Phys. Rev. Lett.}\ }\textbf {\bibinfo {volume} {99}},\
  \bibinfo {pages} {187002} (\bibinfo {year} {2007})}\BibitemShut {NoStop}%
\bibitem [{\citenamefont {Agosta}\ \emph {et~al.}(2017)\citenamefont {Agosta},
  \citenamefont {Fortune}, \citenamefont {Hannahs}, \citenamefont {Gu},
  \citenamefont {Liang}, \citenamefont {Park},\ and\ \citenamefont
  {Schleuter}}]{OrganicSpecificHeat}%
  \BibitemOpen
  \bibfield  {author} {\bibinfo {author} {\bibfnamefont {C.~C.}\ \bibnamefont
  {Agosta}}, \bibinfo {author} {\bibfnamefont {N.~A.}\ \bibnamefont {Fortune}},
  \bibinfo {author} {\bibfnamefont {S.~T.}\ \bibnamefont {Hannahs}}, \bibinfo
  {author} {\bibfnamefont {S.}~\bibnamefont {Gu}}, \bibinfo {author}
  {\bibfnamefont {L.}~\bibnamefont {Liang}}, \bibinfo {author} {\bibfnamefont
  {J.-H.}\ \bibnamefont {Park}},\ and\ \bibinfo {author} {\bibfnamefont
  {J.~A.}\ \bibnamefont {Schleuter}},\ }\bibfield  {title} {\bibinfo {title}
  {{Calorimetric Measurements of Magnetic-Field-Induced Inhomogeneous
  Superconductivity Above the Paramagnetic Limit}},\ }\href
  {https://doi.org/10.1103/PhysRevLett.118.267001} {\bibfield  {journal}
  {\bibinfo  {journal} {Phys. Rev. Lett.}\ }\textbf {\bibinfo {volume} {118}},\
  \bibinfo {pages} {267001} (\bibinfo {year} {2017})}\BibitemShut {NoStop}%
\bibitem [{\citenamefont {Cho}\ \emph {et~al.}(2017)\citenamefont {Cho},
  \citenamefont {Yang}, \citenamefont {Yuan}, \citenamefont {Shen},
  \citenamefont {Wolf},\ and\ \citenamefont {Lortz}}]{FFLOKFe2As2}%
  \BibitemOpen
  \bibfield  {author} {\bibinfo {author} {\bibfnamefont {C.-w.}\ \bibnamefont
  {Cho}}, \bibinfo {author} {\bibfnamefont {J.~H.}\ \bibnamefont {Yang}},
  \bibinfo {author} {\bibfnamefont {N.~F.~Q.}\ \bibnamefont {Yuan}}, \bibinfo
  {author} {\bibfnamefont {J.}~\bibnamefont {Shen}}, \bibinfo {author}
  {\bibfnamefont {T.}~\bibnamefont {Wolf}},\ and\ \bibinfo {author}
  {\bibfnamefont {R.}~\bibnamefont {Lortz}},\ }\bibfield  {title} {\bibinfo
  {title} {{Thermodynamic Evidence for the Fulde-Ferrell-Larkin-Ovchinnikov
  State in the ${\mathrm{KFe}}_{2}{\mathrm{As}}_{2}$ Superconductor}},\ }\href
  {https://doi.org/10.1103/PhysRevLett.119.217002} {\bibfield  {journal}
  {\bibinfo  {journal} {Phys. Rev. Lett.}\ }\textbf {\bibinfo {volume} {119}},\
  \bibinfo {pages} {217002} (\bibinfo {year} {2017})}\BibitemShut {NoStop}%
\bibitem [{\citenamefont {Wan}\ \emph {et~al.}(2023)\citenamefont {Wan},
  \citenamefont {Zheliuk}, \citenamefont {Yuan}, \citenamefont {Peng},
  \citenamefont {Zhang}, \citenamefont {Liang}, \citenamefont {Zeitler},
  \citenamefont {Wiedmann}, \citenamefont {Hussey}, \citenamefont {Palstra},\
  and\ \citenamefont {Ye}}]{Wan2023orbitalFFLO}%
  \BibitemOpen
  \bibfield  {author} {\bibinfo {author} {\bibfnamefont {P.}~\bibnamefont
  {Wan}}, \bibinfo {author} {\bibfnamefont {O.}~\bibnamefont {Zheliuk}},
  \bibinfo {author} {\bibfnamefont {N.~F.~Q.}\ \bibnamefont {Yuan}}, \bibinfo
  {author} {\bibfnamefont {X.}~\bibnamefont {Peng}}, \bibinfo {author}
  {\bibfnamefont {L.}~\bibnamefont {Zhang}}, \bibinfo {author} {\bibfnamefont
  {M.}~\bibnamefont {Liang}}, \bibinfo {author} {\bibfnamefont
  {U.}~\bibnamefont {Zeitler}}, \bibinfo {author} {\bibfnamefont
  {S.}~\bibnamefont {Wiedmann}}, \bibinfo {author} {\bibfnamefont {N.~E.}\
  \bibnamefont {Hussey}}, \bibinfo {author} {\bibfnamefont {T.~T.~M.}\
  \bibnamefont {Palstra}},\ and\ \bibinfo {author} {\bibfnamefont
  {J.}~\bibnamefont {Ye}},\ }\bibfield  {title} {\bibinfo {title} {{Orbital
  Fulde--Ferrell--Larkin--Ovchinnikov state in an Ising superconductor}},\
  }\href {https://doi.org/10.1038/s41586-023-05967-z} {\bibfield  {journal}
  {\bibinfo  {journal} {Nature}\ }\textbf {\bibinfo {volume} {619}},\ \bibinfo
  {pages} {46} (\bibinfo {year} {2023})}\BibitemShut {NoStop}%
\bibitem [{\citenamefont {Itahashi}\ \emph {et~al.}(2025)\citenamefont
  {Itahashi}, \citenamefont {Nohara}, \citenamefont {Chazono}, \citenamefont
  {Matsuoka}, \citenamefont {Arioka}, \citenamefont {Nomoto}, \citenamefont
  {Kohama}, \citenamefont {Yanase}, \citenamefont {Iwasa},\ and\ \citenamefont
  {Kobayashi}}]{Itahashi2025Misfit}%
  \BibitemOpen
  \bibfield  {author} {\bibinfo {author} {\bibfnamefont {Y.~M.}\ \bibnamefont
  {Itahashi}}, \bibinfo {author} {\bibfnamefont {Y.}~\bibnamefont {Nohara}},
  \bibinfo {author} {\bibfnamefont {M.}~\bibnamefont {Chazono}}, \bibinfo
  {author} {\bibfnamefont {H.}~\bibnamefont {Matsuoka}}, \bibinfo {author}
  {\bibfnamefont {K.}~\bibnamefont {Arioka}}, \bibinfo {author} {\bibfnamefont
  {T.}~\bibnamefont {Nomoto}}, \bibinfo {author} {\bibfnamefont
  {Y.}~\bibnamefont {Kohama}}, \bibinfo {author} {\bibfnamefont
  {Y.}~\bibnamefont {Yanase}}, \bibinfo {author} {\bibfnamefont
  {Y.}~\bibnamefont {Iwasa}},\ and\ \bibinfo {author} {\bibfnamefont
  {K.}~\bibnamefont {Kobayashi}},\ }\bibfield  {title} {\bibinfo {title}
  {{Misfit layered superconductor (PbSe)$_{1.14}$(NbSe$_2$)$_3$ with possible
  layer-selective FFLO state}},\ }\href
  {https://doi.org/10.1038/s41467-025-62297-6} {\bibfield  {journal} {\bibinfo
  {journal} {Nature Communications}\ }\textbf {\bibinfo {volume} {16}},\
  \bibinfo {pages} {7022} (\bibinfo {year} {2025})}\BibitemShut {NoStop}%
\bibitem [{\citenamefont {White}(1992)}]{DMRGprl}%
  \BibitemOpen
  \bibfield  {author} {\bibinfo {author} {\bibfnamefont {S.~R.}\ \bibnamefont
  {White}},\ }\bibfield  {title} {\bibinfo {title} {{Density matrix formulation
  for quantum renormalization groups}},\ }\href
  {https://doi.org/10.1103/PhysRevLett.69.2863} {\bibfield  {journal} {\bibinfo
   {journal} {Phys. Rev. Lett.}\ }\textbf {\bibinfo {volume} {69}},\ \bibinfo
  {pages} {2863} (\bibinfo {year} {1992})}\BibitemShut {NoStop}%
\bibitem [{\citenamefont {Schollwöck}(2011)}]{DMRGmps}%
  \BibitemOpen
  \bibfield  {author} {\bibinfo {author} {\bibfnamefont {U.}~\bibnamefont
  {Schollwöck}},\ }\bibfield  {title} {\bibinfo {title} {{The density-matrix
  renormalization group in the age of matrix product states}},\ }\href
  {https://doi.org/https://doi.org/10.1016/j.aop.2010.09.012} {\bibfield
  {journal} {\bibinfo  {journal} {Annals of Physics}\ }\textbf {\bibinfo
  {volume} {326}},\ \bibinfo {pages} {96} (\bibinfo {year} {2011})},\ \bibinfo
  {note} {january 2011 Special Issue}\BibitemShut {NoStop}%
\bibitem [{\citenamefont {Li}\ \emph {et~al.}(2023)\citenamefont {Li},
  \citenamefont {Gao}, \citenamefont {He}, \citenamefont {Qi}, \citenamefont
  {Chen},\ and\ \citenamefont {Li}}]{tanTRG2023}%
  \BibitemOpen
  \bibfield  {author} {\bibinfo {author} {\bibfnamefont {Q.}~\bibnamefont
  {Li}}, \bibinfo {author} {\bibfnamefont {Y.}~\bibnamefont {Gao}}, \bibinfo
  {author} {\bibfnamefont {Y.-Y.}\ \bibnamefont {He}}, \bibinfo {author}
  {\bibfnamefont {Y.}~\bibnamefont {Qi}}, \bibinfo {author} {\bibfnamefont
  {B.-B.}\ \bibnamefont {Chen}},\ and\ \bibinfo {author} {\bibfnamefont
  {W.}~\bibnamefont {Li}},\ }\bibfield  {title} {\bibinfo {title} {{Tangent
  Space Approach for Thermal Tensor Network Simulations of the {2D} {Hubbard}
  Model}},\ }\href {https://doi.org/10.1103/PhysRevLett.130.226502} {\bibfield
  {journal} {\bibinfo  {journal} {Phys. Rev. Lett.}\ }\textbf {\bibinfo
  {volume} {130}},\ \bibinfo {pages} {226502} (\bibinfo {year}
  {2023})}\BibitemShut {NoStop}%
\bibitem [{\citenamefont {Li}(2025)}]{Li_FiniteMPS_jl_2025}%
  \BibitemOpen
  \bibfield  {author} {\bibinfo {author} {\bibfnamefont {Q.}~\bibnamefont
  {Li}},\ }\href {https://doi.org/10.5281/zenodo.14615184} {\bibinfo {title}
  {{FiniteMPS.jl}}} (\bibinfo {year} {2025})\BibitemShut {NoStop}%
\bibitem [{\citenamefont {Weichselbaum}(2012)}]{QSpace}%
  \BibitemOpen
  \bibfield  {author} {\bibinfo {author} {\bibfnamefont {A.}~\bibnamefont
  {Weichselbaum}},\ }\bibfield  {title} {\bibinfo {title} {{Non-abelian
  symmetries in tensor networks: A quantum symmetry space approach}},\ }\href
  {https://doi.org/https://doi.org/10.1016/j.aop.2012.07.009} {\bibfield
  {journal} {\bibinfo  {journal} {Annals of Physics}\ }\textbf {\bibinfo
  {volume} {327}},\ \bibinfo {pages} {2972} (\bibinfo {year}
  {2012})}\BibitemShut {NoStop}%
\bibitem [{\citenamefont {Devos}\ and\ \citenamefont
  {Haegeman}(2025)}]{TensorKit}%
  \BibitemOpen
  \bibfield  {author} {\bibinfo {author} {\bibfnamefont {L.}~\bibnamefont
  {Devos}}\ and\ \bibinfo {author} {\bibfnamefont {J.}~\bibnamefont
  {Haegeman}},\ }\href@noop {} {\bibinfo {title} {{TensorKit.jl: A Julia
  package for large-scale tensor computations, with a hint of category
  theory}}} (\bibinfo {year} {2025}),\ \Eprint
  {https://arxiv.org/abs/2508.10076} {arXiv:2508.10076 [cs.MS]} \BibitemShut
  {NoStop}%
\bibitem [{SM()}]{SM}%
  \BibitemOpen
  \href@noop {} {\bibinfo {title} {{In Supplemental Material
  Sec.$\mathrm{\uppercase\expandafter{\romannumeral1}}$, we show the
  convergence check of DMRG and tanTRG calculations. In
  Sec.$\mathrm{\uppercase\expandafter{\romannumeral2}}$, we provide details of
  ground-state spin correlations and single-particle Green's functions. In
  Sec.$\mathrm{\uppercase\expandafter{\romannumeral4}}$, we show the real space
  charge and spin density profiles. In
  Sec.$\mathrm{\uppercase\expandafter{\romannumeral4}}$, we present DMRG
  results on width-$4$ cylindrical systems.}}}\BibitemShut {Stop}%
\bibitem [{\citenamefont {Li}\ \emph {et~al.}(2025)\citenamefont {Li},
  \citenamefont {Qu}, \citenamefont {Chen}, \citenamefont {Shi},\ and\
  \citenamefont {Li}}]{li2025fixNtanTRG}%
  \BibitemOpen
  \bibfield  {author} {\bibinfo {author} {\bibfnamefont {Q.}~\bibnamefont
  {Li}}, \bibinfo {author} {\bibfnamefont {D.-W.}\ \bibnamefont {Qu}}, \bibinfo
  {author} {\bibfnamefont {B.-B.}\ \bibnamefont {Chen}}, \bibinfo {author}
  {\bibfnamefont {T.}~\bibnamefont {Shi}},\ and\ \bibinfo {author}
  {\bibfnamefont {W.}~\bibnamefont {Li}},\ }\href@noop {} {\bibinfo {title}
  {{Thermal Tensor Network Simulations of Fermions with a Fixed Filling}}}
  (\bibinfo {year} {2025}),\ \Eprint {https://arxiv.org/abs/2511.07303}
  {arXiv:2511.07303 [cond-mat.str-el]} \BibitemShut {NoStop}%
\bibitem [{\citenamefont {Lu}\ \emph {et~al.}(2023)\citenamefont {Lu},
  \citenamefont {Qu}, \citenamefont {Qi}, \citenamefont {Li},\ and\
  \citenamefont {Gong}}]{LuXintwoleg}%
  \BibitemOpen
  \bibfield  {author} {\bibinfo {author} {\bibfnamefont {X.}~\bibnamefont
  {Lu}}, \bibinfo {author} {\bibfnamefont {D.-W.}\ \bibnamefont {Qu}}, \bibinfo
  {author} {\bibfnamefont {Y.}~\bibnamefont {Qi}}, \bibinfo {author}
  {\bibfnamefont {W.}~\bibnamefont {Li}},\ and\ \bibinfo {author}
  {\bibfnamefont {S.-S.}\ \bibnamefont {Gong}},\ }\bibfield  {title} {\bibinfo
  {title} {{Ground-state phase diagram of the extended two-leg
  $t\text{\ensuremath{-}}J$ ladder}},\ }\href
  {https://doi.org/10.1103/PhysRevB.107.125114} {\bibfield  {journal} {\bibinfo
   {journal} {Phys. Rev. B}\ }\textbf {\bibinfo {volume} {107}},\ \bibinfo
  {pages} {125114} (\bibinfo {year} {2023})}\BibitemShut {NoStop}%
\bibitem [{\citenamefont {Oshikawa}\ \emph {et~al.}(1997)\citenamefont
  {Oshikawa}, \citenamefont {Yamanaka},\ and\ \citenamefont {Affleck}}]{OYA}%
  \BibitemOpen
  \bibfield  {author} {\bibinfo {author} {\bibfnamefont {M.}~\bibnamefont
  {Oshikawa}}, \bibinfo {author} {\bibfnamefont {M.}~\bibnamefont {Yamanaka}},\
  and\ \bibinfo {author} {\bibfnamefont {I.}~\bibnamefont {Affleck}},\
  }\bibfield  {title} {\bibinfo {title} {{Magnetization Plateaus in Spin
  Chains: ''Haldane Gap'' for Half-Integer Spins}},\ }\href
  {https://doi.org/10.1103/PhysRevLett.78.1984} {\bibfield  {journal} {\bibinfo
   {journal} {Phys. Rev. Lett.}\ }\textbf {\bibinfo {volume} {78}},\ \bibinfo
  {pages} {1984} (\bibinfo {year} {1997})}\BibitemShut {NoStop}%
\bibitem [{\citenamefont {Yamanaka}\ \emph {et~al.}(1997)\citenamefont
  {Yamanaka}, \citenamefont {Oshikawa},\ and\ \citenamefont {Affleck}}]{OYA1D}%
  \BibitemOpen
  \bibfield  {author} {\bibinfo {author} {\bibfnamefont {M.}~\bibnamefont
  {Yamanaka}}, \bibinfo {author} {\bibfnamefont {M.}~\bibnamefont {Oshikawa}},\
  and\ \bibinfo {author} {\bibfnamefont {I.}~\bibnamefont {Affleck}},\
  }\bibfield  {title} {\bibinfo {title} {{Nonperturbative Approach to
  Luttinger's Theorem in One Dimension}},\ }\href
  {https://doi.org/10.1103/PhysRevLett.79.1110} {\bibfield  {journal} {\bibinfo
   {journal} {Phys. Rev. Lett.}\ }\textbf {\bibinfo {volume} {79}},\ \bibinfo
  {pages} {1110} (\bibinfo {year} {1997})}\BibitemShut {NoStop}%
\bibitem [{\citenamefont {Pruschke}\ and\ \citenamefont
  {Shiba}(1992{\natexlab{b}})}]{Pruschke1992superconducting}%
  \BibitemOpen
  \bibfield  {author} {\bibinfo {author} {\bibfnamefont {T.}~\bibnamefont
  {Pruschke}}\ and\ \bibinfo {author} {\bibfnamefont {H.}~\bibnamefont
  {Shiba}},\ }\bibfield  {title} {\bibinfo {title} {{Superconducting
  correlations in the one-dimensional $t$-$J$ model}},\ }\href
  {https://doi.org/10.1103/PhysRevB.46.356} {\bibfield  {journal} {\bibinfo
  {journal} {Phys. Rev. B}\ }\textbf {\bibinfo {volume} {46}},\ \bibinfo
  {pages} {356} (\bibinfo {year} {1992}{\natexlab{b}})}\BibitemShut {NoStop}%
\bibitem [{\citenamefont {Wietek}(2022)}]{Wietek2022}%
  \BibitemOpen
  \bibfield  {author} {\bibinfo {author} {\bibfnamefont {A.}~\bibnamefont
  {Wietek}},\ }\bibfield  {title} {\bibinfo {title} {{Fragmented Cooper Pair
  Condensation in Striped Superconductors}},\ }\href
  {https://doi.org/10.1103/PhysRevLett.129.177001} {\bibfield  {journal}
  {\bibinfo  {journal} {Phys. Rev. Lett.}\ }\textbf {\bibinfo {volume} {129}},\
  \bibinfo {pages} {177001} (\bibinfo {year} {2022})}\BibitemShut {NoStop}%
\bibitem [{\citenamefont {Leggett}(1970)}]{Leggett1970SS}%
  \BibitemOpen
  \bibfield  {author} {\bibinfo {author} {\bibfnamefont {A.~J.}\ \bibnamefont
  {Leggett}},\ }\bibfield  {title} {\bibinfo {title} {{Can a Solid Be
  "Superfluid"?}},\ }\href {https://doi.org/10.1103/PhysRevLett.25.1543}
  {\bibfield  {journal} {\bibinfo  {journal} {Phys. Rev. Lett.}\ }\textbf
  {\bibinfo {volume} {25}},\ \bibinfo {pages} {1543} (\bibinfo {year}
  {1970})}\BibitemShut {NoStop}%
\bibitem [{\citenamefont {Xiang}\ \emph {et~al.}(2024)\citenamefont {Xiang},
  \citenamefont {Zhang}, \citenamefont {Gao}, \citenamefont {Schmidt},
  \citenamefont {Schmalzl}, \citenamefont {Wang}, \citenamefont {Li},
  \citenamefont {Xi}, \citenamefont {Liu}, \citenamefont {Jin}, \citenamefont
  {Li}, \citenamefont {Shen}, \citenamefont {Chen}, \citenamefont {Qi},
  \citenamefont {Wan}, \citenamefont {Jin}, \citenamefont {Li}, \citenamefont
  {Sun},\ and\ \citenamefont {Su}}]{Xiang2024SS}%
  \BibitemOpen
  \bibfield  {author} {\bibinfo {author} {\bibfnamefont {J.}~\bibnamefont
  {Xiang}}, \bibinfo {author} {\bibfnamefont {C.}~\bibnamefont {Zhang}},
  \bibinfo {author} {\bibfnamefont {Y.}~\bibnamefont {Gao}}, \bibinfo {author}
  {\bibfnamefont {W.}~\bibnamefont {Schmidt}}, \bibinfo {author} {\bibfnamefont
  {K.}~\bibnamefont {Schmalzl}}, \bibinfo {author} {\bibfnamefont {C.-W.}\
  \bibnamefont {Wang}}, \bibinfo {author} {\bibfnamefont {B.}~\bibnamefont
  {Li}}, \bibinfo {author} {\bibfnamefont {N.}~\bibnamefont {Xi}}, \bibinfo
  {author} {\bibfnamefont {X.-Y.}\ \bibnamefont {Liu}}, \bibinfo {author}
  {\bibfnamefont {H.}~\bibnamefont {Jin}}, \bibinfo {author} {\bibfnamefont
  {G.}~\bibnamefont {Li}}, \bibinfo {author} {\bibfnamefont {J.}~\bibnamefont
  {Shen}}, \bibinfo {author} {\bibfnamefont {Z.}~\bibnamefont {Chen}}, \bibinfo
  {author} {\bibfnamefont {Y.}~\bibnamefont {Qi}}, \bibinfo {author}
  {\bibfnamefont {Y.}~\bibnamefont {Wan}}, \bibinfo {author} {\bibfnamefont
  {W.}~\bibnamefont {Jin}}, \bibinfo {author} {\bibfnamefont {W.}~\bibnamefont
  {Li}}, \bibinfo {author} {\bibfnamefont {P.}~\bibnamefont {Sun}},\ and\
  \bibinfo {author} {\bibfnamefont {G.}~\bibnamefont {Su}},\ }\bibfield
  {title} {\bibinfo {title} {{Giant magnetocaloric effect in spin supersolid
  candidate Na$_2$BaCo(PO$_4$)$_2$}},\ }\href
  {https://doi.org/10.1038/s41586-023-06885-w} {\bibfield  {journal} {\bibinfo
  {journal} {Nature}\ }\textbf {\bibinfo {volume} {625}},\ \bibinfo {pages}
  {270} (\bibinfo {year} {2024})}\BibitemShut {NoStop}%
\bibitem [{\citenamefont {Recati}\ and\ \citenamefont
  {Stringari}(2023)}]{Recati2023SS}%
  \BibitemOpen
  \bibfield  {author} {\bibinfo {author} {\bibfnamefont {A.}~\bibnamefont
  {Recati}}\ and\ \bibinfo {author} {\bibfnamefont {S.}~\bibnamefont
  {Stringari}},\ }\bibfield  {title} {\bibinfo {title} {{Supersolidity in
  ultracold dipolar gases}},\ }\href
  {https://doi.org/10.1038/s42254-023-00648-2} {\bibfield  {journal} {\bibinfo
  {journal} {Nature Reviews Physics}\ }\textbf {\bibinfo {volume} {5}},\
  \bibinfo {pages} {735} (\bibinfo {year} {2023})}\BibitemShut {NoStop}%
\bibitem [{\citenamefont {Xie}\ and\ \citenamefont
  {Nagaosa}(2025)}]{xie2025SS}%
  \BibitemOpen
  \bibfield  {author} {\bibinfo {author} {\bibfnamefont {Y.-M.}\ \bibnamefont
  {Xie}}\ and\ \bibinfo {author} {\bibfnamefont {N.}~\bibnamefont {Nagaosa}},\
  }\bibfield  {title} {\bibinfo {title} {{Quasi-one-dimensional supersolids in
  Luther-Emery liquids}},\ }\href {https://doi.org/10.1103/h3l5-njrd}
  {\bibfield  {journal} {\bibinfo  {journal} {Phys. Rev. Res.}\ }\textbf
  {\bibinfo {volume} {7}},\ \bibinfo {pages} {023302} (\bibinfo {year}
  {2025})}\BibitemShut {NoStop}%
\bibitem [{\citenamefont {Meng}\ \emph {et~al.}(2024)\citenamefont {Meng},
  \citenamefont {Zhang}, \citenamefont {Fernandes}, \citenamefont {Ma},\ and\
  \citenamefont {Scalettar}}]{Meng2025SS}%
  \BibitemOpen
  \bibfield  {author} {\bibinfo {author} {\bibfnamefont {J.}~\bibnamefont
  {Meng}}, \bibinfo {author} {\bibfnamefont {Y.}~\bibnamefont {Zhang}},
  \bibinfo {author} {\bibfnamefont {R.~M.}\ \bibnamefont {Fernandes}}, \bibinfo
  {author} {\bibfnamefont {T.}~\bibnamefont {Ma}},\ and\ \bibinfo {author}
  {\bibfnamefont {R.~T.}\ \bibnamefont {Scalettar}},\ }\bibfield  {title}
  {\bibinfo {title} {{Supersolid phase in the diluted Holstein model}},\ }\href
  {https://doi.org/10.1103/PhysRevB.110.L220506} {\bibfield  {journal}
  {\bibinfo  {journal} {Phys. Rev. B}\ }\textbf {\bibinfo {volume} {110}},\
  \bibinfo {pages} {L220506} (\bibinfo {year} {2024})}\BibitemShut {NoStop}%
\bibitem [{\citenamefont {Huscroft}\ \emph {et~al.}(2001)\citenamefont
  {Huscroft}, \citenamefont {Jarrell}, \citenamefont {Maier}, \citenamefont
  {Moukouri},\ and\ \citenamefont {Tahvildarzadeh}}]{pseudogap2001}%
  \BibitemOpen
  \bibfield  {author} {\bibinfo {author} {\bibfnamefont {C.}~\bibnamefont
  {Huscroft}}, \bibinfo {author} {\bibfnamefont {M.}~\bibnamefont {Jarrell}},
  \bibinfo {author} {\bibfnamefont {T.}~\bibnamefont {Maier}}, \bibinfo
  {author} {\bibfnamefont {S.}~\bibnamefont {Moukouri}},\ and\ \bibinfo
  {author} {\bibfnamefont {A.~N.}\ \bibnamefont {Tahvildarzadeh}},\ }\bibfield
  {title} {\bibinfo {title} {{Pseudogaps in the 2D Hubbard Model}},\ }\href
  {https://doi.org/10.1103/PhysRevLett.86.139} {\bibfield  {journal} {\bibinfo
  {journal} {Phys. Rev. Lett.}\ }\textbf {\bibinfo {volume} {86}},\ \bibinfo
  {pages} {139} (\bibinfo {year} {2001})}\BibitemShut {NoStop}%
\bibitem [{\citenamefont {Gull}\ \emph {et~al.}(2013)\citenamefont {Gull},
  \citenamefont {Parcollet},\ and\ \citenamefont {Millis}}]{pseudogap2013}%
  \BibitemOpen
  \bibfield  {author} {\bibinfo {author} {\bibfnamefont {E.}~\bibnamefont
  {Gull}}, \bibinfo {author} {\bibfnamefont {O.}~\bibnamefont {Parcollet}},\
  and\ \bibinfo {author} {\bibfnamefont {A.~J.}\ \bibnamefont {Millis}},\
  }\bibfield  {title} {\bibinfo {title} {{Superconductivity and the Pseudogap
  in the Two-Dimensional Hubbard Model}},\ }\href
  {https://doi.org/10.1103/PhysRevLett.110.216405} {\bibfield  {journal}
  {\bibinfo  {journal} {Phys. Rev. Lett.}\ }\textbf {\bibinfo {volume} {110}},\
  \bibinfo {pages} {216405} (\bibinfo {year} {2013})}\BibitemShut {NoStop}%
\bibitem [{\citenamefont {Rubtsov}\ \emph {et~al.}(2009)\citenamefont
  {Rubtsov}, \citenamefont {Katsnelson}, \citenamefont {Lichtenstein},\ and\
  \citenamefont {Georges}}]{pseudogap2009}%
  \BibitemOpen
  \bibfield  {author} {\bibinfo {author} {\bibfnamefont {A.~N.}\ \bibnamefont
  {Rubtsov}}, \bibinfo {author} {\bibfnamefont {M.~I.}\ \bibnamefont
  {Katsnelson}}, \bibinfo {author} {\bibfnamefont {A.~I.}\ \bibnamefont
  {Lichtenstein}},\ and\ \bibinfo {author} {\bibfnamefont {A.}~\bibnamefont
  {Georges}},\ }\bibfield  {title} {\bibinfo {title} {{Dual fermion approach to
  the two-dimensional Hubbard model: Antiferromagnetic fluctuations and Fermi
  arcs}},\ }\href {https://doi.org/10.1103/PhysRevB.79.045133} {\bibfield
  {journal} {\bibinfo  {journal} {Phys. Rev. B}\ }\textbf {\bibinfo {volume}
  {79}},\ \bibinfo {pages} {045133} (\bibinfo {year} {2009})}\BibitemShut
  {NoStop}%
\bibitem [{\citenamefont {Jiang}\ \emph {et~al.}(2022)\citenamefont {Jiang},
  \citenamefont {Liu}, \citenamefont {Klein}, \citenamefont {Wang},
  \citenamefont {Sun}, \citenamefont {Chubukov},\ and\ \citenamefont
  {Meng}}]{Jiang2022ITP}%
  \BibitemOpen
  \bibfield  {author} {\bibinfo {author} {\bibfnamefont {W.}~\bibnamefont
  {Jiang}}, \bibinfo {author} {\bibfnamefont {Y.}~\bibnamefont {Liu}}, \bibinfo
  {author} {\bibfnamefont {A.}~\bibnamefont {Klein}}, \bibinfo {author}
  {\bibfnamefont {Y.}~\bibnamefont {Wang}}, \bibinfo {author} {\bibfnamefont
  {K.}~\bibnamefont {Sun}}, \bibinfo {author} {\bibfnamefont {A.~V.}\
  \bibnamefont {Chubukov}},\ and\ \bibinfo {author} {\bibfnamefont {Z.~Y.}\
  \bibnamefont {Meng}},\ }\bibfield  {title} {\bibinfo {title} {{Monte Carlo
  study of the pseudogap and superconductivity emerging from quantum magnetic
  fluctuations}},\ }\href {https://doi.org/10.1038/s41467-022-30302-x}
  {\bibfield  {journal} {\bibinfo  {journal} {Nature Communications}\ }\textbf
  {\bibinfo {volume} {13}},\ \bibinfo {pages} {2655} (\bibinfo {year}
  {2022})}\BibitemShut {NoStop}%
\bibitem [{\citenamefont {Lederer}\ \emph {et~al.}(2017)\citenamefont
  {Lederer}, \citenamefont {Schattner}, \citenamefont {Berg},\ and\
  \citenamefont {Kivelson}}]{Samuel2017ITP}%
  \BibitemOpen
  \bibfield  {author} {\bibinfo {author} {\bibfnamefont {S.}~\bibnamefont
  {Lederer}}, \bibinfo {author} {\bibfnamefont {Y.}~\bibnamefont {Schattner}},
  \bibinfo {author} {\bibfnamefont {E.}~\bibnamefont {Berg}},\ and\ \bibinfo
  {author} {\bibfnamefont {S.~A.}\ \bibnamefont {Kivelson}},\ }\bibfield
  {title} {\bibinfo {title} {{Superconductivity and non-Fermi liquid behavior
  near a nematic quantum critical point}},\ }\href
  {https://doi.org/10.1073/pnas.1620651114} {\bibfield  {journal} {\bibinfo
  {journal} {Proceedings of the National Academy of Sciences}\ }\textbf
  {\bibinfo {volume} {114}},\ \bibinfo {pages} {4905} (\bibinfo {year}
  {2017})}\BibitemShut {NoStop}%
\bibitem [{\citenamefont {Xu}\ \emph {et~al.}(2022)\citenamefont {Xu},
  \citenamefont {Shi}, \citenamefont {Vitali}, \citenamefont {Qin},\ and\
  \citenamefont {Zhang}}]{ZhangShiweiStripe2022}%
  \BibitemOpen
  \bibfield  {author} {\bibinfo {author} {\bibfnamefont {H.}~\bibnamefont
  {Xu}}, \bibinfo {author} {\bibfnamefont {H.}~\bibnamefont {Shi}}, \bibinfo
  {author} {\bibfnamefont {E.}~\bibnamefont {Vitali}}, \bibinfo {author}
  {\bibfnamefont {M.}~\bibnamefont {Qin}},\ and\ \bibinfo {author}
  {\bibfnamefont {S.}~\bibnamefont {Zhang}},\ }\bibfield  {title} {\bibinfo
  {title} {{Stripes and spin-density waves in the doped two-dimensional Hubbard
  model: Ground state phase diagram}},\ }\href
  {https://doi.org/10.1103/PhysRevResearch.4.013239} {\bibfield  {journal}
  {\bibinfo  {journal} {Phys. Rev. Res.}\ }\textbf {\bibinfo {volume} {4}},\
  \bibinfo {pages} {013239} (\bibinfo {year} {2022})}\BibitemShut {NoStop}%
\bibitem [{\citenamefont {Lyons}\ \emph {et~al.}(1988)\citenamefont {Lyons},
  \citenamefont {Fleury}, \citenamefont {Remeika}, \citenamefont {Cooper},\
  and\ \citenamefont {Negran}}]{lyons1988dynamics}%
  \BibitemOpen
  \bibfield  {author} {\bibinfo {author} {\bibfnamefont {K.~B.}\ \bibnamefont
  {Lyons}}, \bibinfo {author} {\bibfnamefont {P.~A.}\ \bibnamefont {Fleury}},
  \bibinfo {author} {\bibfnamefont {J.~P.}\ \bibnamefont {Remeika}}, \bibinfo
  {author} {\bibfnamefont {A.~S.}\ \bibnamefont {Cooper}},\ and\ \bibinfo
  {author} {\bibfnamefont {T.~J.}\ \bibnamefont {Negran}},\ }\bibfield  {title}
  {\bibinfo {title} {Dynamics of spin fluctuations in lanthanum cuprate},\
  }\href {https://doi.org/10.1103/PhysRevB.37.2353} {\bibfield  {journal}
  {\bibinfo  {journal} {Phys. Rev. B}\ }\textbf {\bibinfo {volume} {37}},\
  \bibinfo {pages} {2353} (\bibinfo {year} {1988})}\BibitemShut {NoStop}%
\bibitem [{\citenamefont {Nakamura}\ \emph {et~al.}(2019)\citenamefont
  {Nakamura}, \citenamefont {Adachi}, \citenamefont {Omori}, \citenamefont
  {Koike},\ and\ \citenamefont {Takeyama}}]{LSCOparallfield}%
  \BibitemOpen
  \bibfield  {author} {\bibinfo {author} {\bibfnamefont {D.}~\bibnamefont
  {Nakamura}}, \bibinfo {author} {\bibfnamefont {T.}~\bibnamefont {Adachi}},
  \bibinfo {author} {\bibfnamefont {K.}~\bibnamefont {Omori}}, \bibinfo
  {author} {\bibfnamefont {Y.}~\bibnamefont {Koike}},\ and\ \bibinfo {author}
  {\bibfnamefont {S.}~\bibnamefont {Takeyama}},\ }\bibfield  {title} {\bibinfo
  {title} {{Pauli-limit upper critical field of high-temperature superconductor
  La$_{1.84}$Sr$_{0.16}$CuO$_4$}},\ }\bibfield  {journal} {\bibinfo  {journal}
  {Scientific Reports}\ }\textbf {\bibinfo {volume} {9}},\ \href
  {https://doi.org/10.1038/s41598-019-52973-1} {10.1038/s41598-019-52973-1}
  (\bibinfo {year} {2019})\BibitemShut {NoStop}%
\bibitem [{\citenamefont {Feng}\ \emph {et~al.}(2025)\citenamefont {Feng},
  \citenamefont {Hartke}, \citenamefont {He}, \citenamefont {Oreg},
  \citenamefont {Turnbaugh}, \citenamefont {Jia}, \citenamefont {Zwierlein},\
  and\ \citenamefont {Zhang}}]{feng2025search}%
  \BibitemOpen
  \bibfield  {author} {\bibinfo {author} {\bibfnamefont {C.}~\bibnamefont
  {Feng}}, \bibinfo {author} {\bibfnamefont {T.}~\bibnamefont {Hartke}},
  \bibinfo {author} {\bibfnamefont {Y.-Y.}\ \bibnamefont {He}}, \bibinfo
  {author} {\bibfnamefont {B.}~\bibnamefont {Oreg}}, \bibinfo {author}
  {\bibfnamefont {C.}~\bibnamefont {Turnbaugh}}, \bibinfo {author}
  {\bibfnamefont {N.}~\bibnamefont {Jia}}, \bibinfo {author} {\bibfnamefont
  {M.}~\bibnamefont {Zwierlein}},\ and\ \bibinfo {author} {\bibfnamefont
  {S.}~\bibnamefont {Zhang}},\ }\href@noop {} {\bibinfo {title} {{In search of
  exotic pairing in the Hubbard model: many-body computation and quantum gas
  microscopy}}} (\bibinfo {year} {2025}),\ \Eprint
  {https://arxiv.org/abs/2509.02688} {arXiv:2509.02688 [cond-mat.quant-gas]}
  \BibitemShut {NoStop}%
\bibitem [{\citenamefont {Kinnunen}\ \emph {et~al.}(2018)\citenamefont
  {Kinnunen}, \citenamefont {Baarsma}, \citenamefont {Martikainen},\ and\
  \citenamefont {Törmä}}]{Kinnunen_2018}%
  \BibitemOpen
  \bibfield  {author} {\bibinfo {author} {\bibfnamefont {J.~J.}\ \bibnamefont
  {Kinnunen}}, \bibinfo {author} {\bibfnamefont {J.~E.}\ \bibnamefont
  {Baarsma}}, \bibinfo {author} {\bibfnamefont {J.-P.}\ \bibnamefont
  {Martikainen}},\ and\ \bibinfo {author} {\bibfnamefont {P.}~\bibnamefont
  {Törmä}},\ }\bibfield  {title} {\bibinfo {title} {{The
  Fulde–Ferrell–Larkin–Ovchinnikov state for ultracold fermions in
  lattice and harmonic potentials: a review}},\ }\href
  {https://doi.org/10.1088/1361-6633/aaa4ad} {\bibfield  {journal} {\bibinfo
  {journal} {Reports on Progress in Physics}\ }\textbf {\bibinfo {volume}
  {81}},\ \bibinfo {pages} {046401} (\bibinfo {year} {2018})}\BibitemShut
  {NoStop}%
\bibitem [{\citenamefont {Revelle}\ \emph {et~al.}(2016)\citenamefont
  {Revelle}, \citenamefont {Fry}, \citenamefont {Olsen},\ and\ \citenamefont
  {Hulet}}]{Revelle20161Dto3D}%
  \BibitemOpen
  \bibfield  {author} {\bibinfo {author} {\bibfnamefont {M.~C.}\ \bibnamefont
  {Revelle}}, \bibinfo {author} {\bibfnamefont {J.~A.}\ \bibnamefont {Fry}},
  \bibinfo {author} {\bibfnamefont {B.~A.}\ \bibnamefont {Olsen}},\ and\
  \bibinfo {author} {\bibfnamefont {R.~G.}\ \bibnamefont {Hulet}},\ }\bibfield
  {title} {\bibinfo {title} {{1D to 3D Crossover of a Spin-Imbalanced Fermi
  Gas}},\ }\href {https://doi.org/10.1103/PhysRevLett.117.235301} {\bibfield
  {journal} {\bibinfo  {journal} {Phys. Rev. Lett.}\ }\textbf {\bibinfo
  {volume} {117}},\ \bibinfo {pages} {235301} (\bibinfo {year}
  {2016})}\BibitemShut {NoStop}%
\bibitem [{\citenamefont {Liao}\ \emph {et~al.}(2010)\citenamefont {Liao},
  \citenamefont {Rittner}, \citenamefont {Paprotta}, \citenamefont {Li},
  \citenamefont {Partridge}, \citenamefont {Hulet}, \citenamefont {Baur},\ and\
  \citenamefont {Mueller}}]{Liao2010SpinImbalance}%
  \BibitemOpen
  \bibfield  {author} {\bibinfo {author} {\bibfnamefont {Y.-a.}\ \bibnamefont
  {Liao}}, \bibinfo {author} {\bibfnamefont {A.~S.~C.}\ \bibnamefont
  {Rittner}}, \bibinfo {author} {\bibfnamefont {T.}~\bibnamefont {Paprotta}},
  \bibinfo {author} {\bibfnamefont {W.}~\bibnamefont {Li}}, \bibinfo {author}
  {\bibfnamefont {G.~B.}\ \bibnamefont {Partridge}}, \bibinfo {author}
  {\bibfnamefont {R.~G.}\ \bibnamefont {Hulet}}, \bibinfo {author}
  {\bibfnamefont {S.~K.}\ \bibnamefont {Baur}},\ and\ \bibinfo {author}
  {\bibfnamefont {E.~J.}\ \bibnamefont {Mueller}},\ }\bibfield  {title}
  {\bibinfo {title} {{Spin-imbalance in a one-dimensional Fermi gas}},\ }\href
  {https://doi.org/10.1038/nature09393} {\bibfield  {journal} {\bibinfo
  {journal} {Nature}\ }\textbf {\bibinfo {volume} {467}},\ \bibinfo {pages}
  {567} (\bibinfo {year} {2010})}\BibitemShut {NoStop}%
\end{thebibliography}%

\clearpage
\newpage
\clearpage
\onecolumngrid
\begin{center}
\section{\textbf{\protect\large{Supplemental Material}}}

\end{center}

\renewcommand{\theequation}{\Alph{section}\arabic{equation}}
\setcounter{section}{0}
\setcounter{figure}{0}
\setcounter{equation}{0}
\renewcommand{\theequation}{S\arabic{equation}}
\renewcommand{\thefigure}{S\arabic{figure}}
\setcounter{secnumdepth}{3}

\section{Convergence Check}
\begin{figure}[htbp]
\centering
\includegraphics[width=0.65\textwidth]{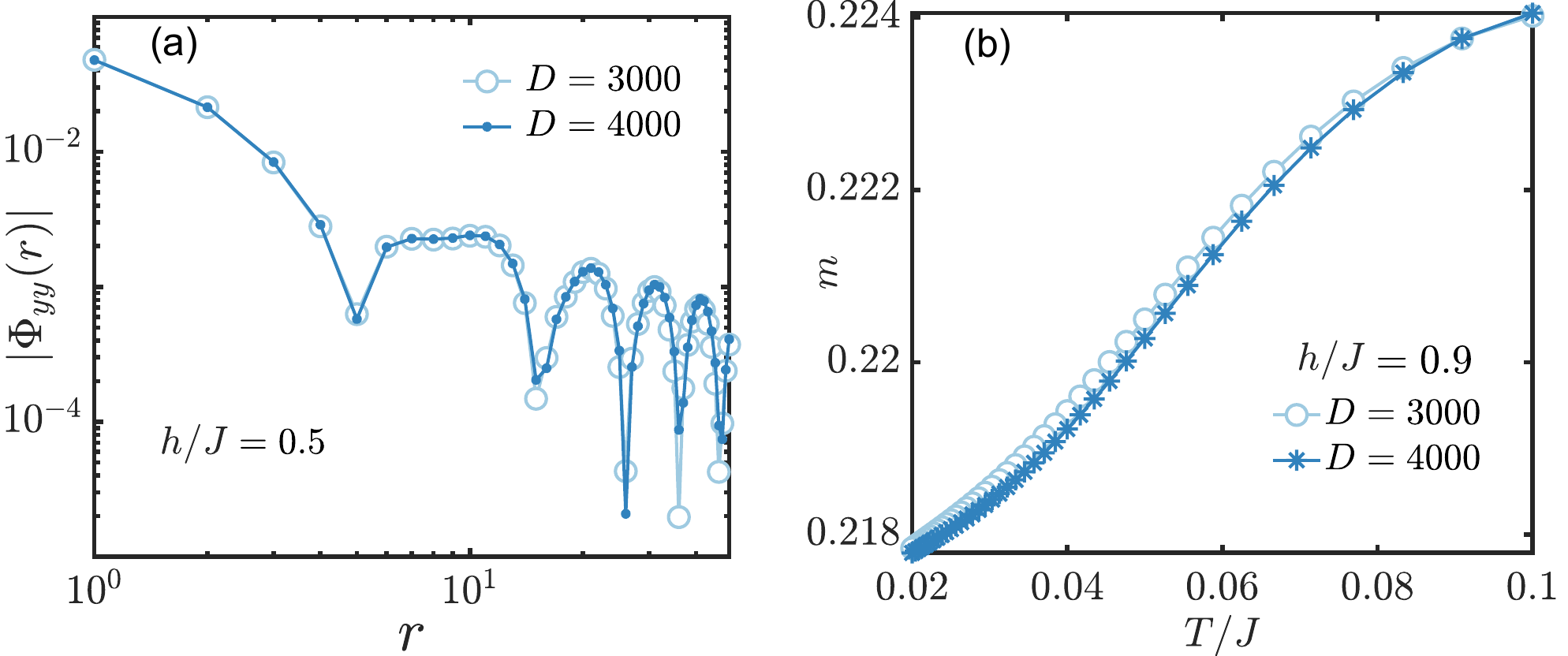}
\caption{(a) Convergence check of ground-state DMRG calculations.
We show the rung pairing correlations at $D = 3000$ and $4000$ DMRG states, obtained from $L_x = 80$, $L_y = 2$
ladder with Zeeman field $h = 0.5 J$.
(b) Convergence check of finite-temperature tanTRG calculations. We show the magnetization $m$ at different 
temperatures, obtained from $L_x = 60$, $L_y = 2$ ladder with $D = 3000$ and $4000$ states. }
\label{FigS1}
\end{figure}
In this section, we show the convergence check of the tensor network calculations.
Figure~\ref{FigS1}(a) illustrates the convergence of DMRG results, with rung pairing correlations 
from $D = 3000$ and $4000$ U(1)$_{\text{charge}} \times$ U(1)$_{\text{spin}}$ states showing 
excellent consistency up to distance $r \sim 20$.
In terms of finite-temperature tanTRG results, 
we show in Fig.~\ref{FigS1}(b) the magnetization $m$ at different temperatures.
The $D = 3000$ and $4000$ results are in good agreement.
Based on our analysis, $D \sim 3000$ states provide sufficient precision for the numerical simulations.

\section{Spin correlations and single-particle Green's function}
\begin{figure}[htbp]
\includegraphics[width=0.7\textwidth]{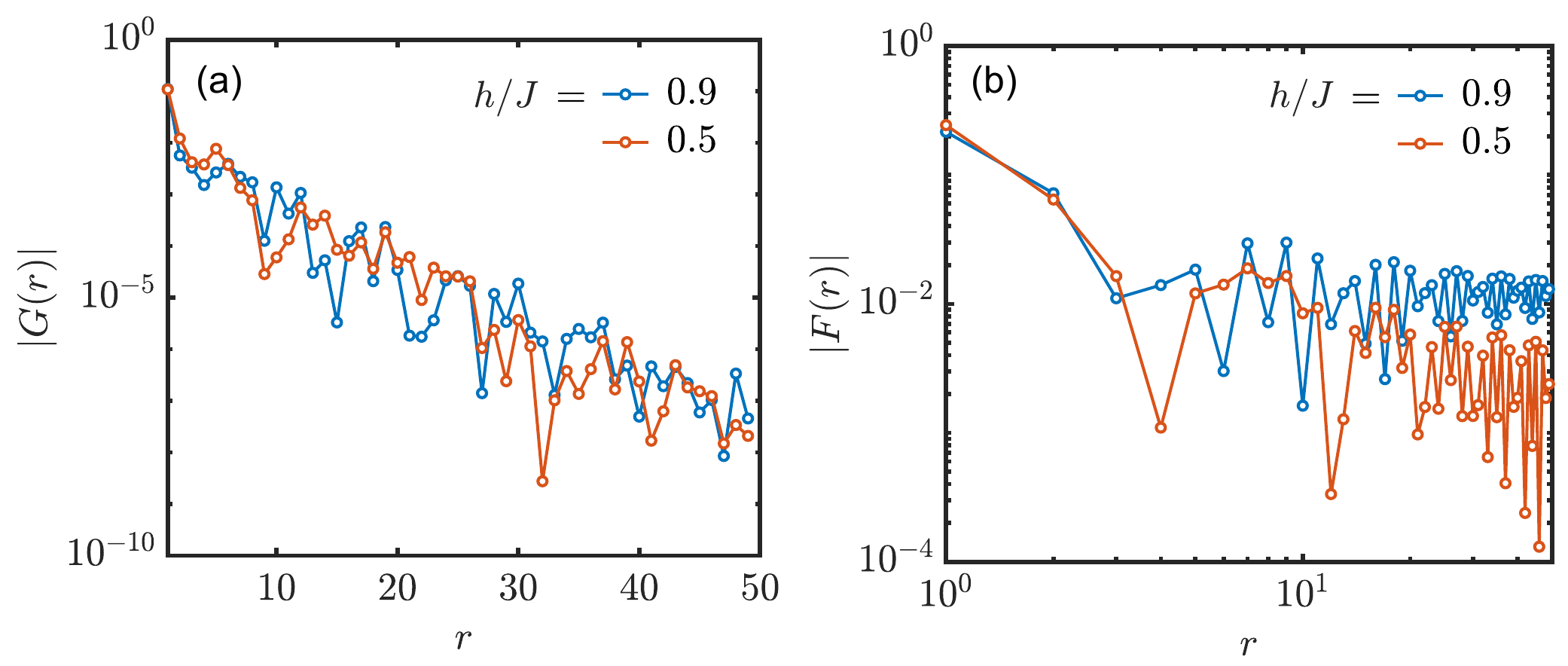}
\caption{
(a) The single-particle Greens's function and (b) the spin correlations of the ground state.
Red and blue colors represent different Zeeman fields $h= 0.5 J$ and $h = 0.9 J$, both exhibiting FFLO superconductivity.}
\label{FigS5}
\end{figure}
In this section, we present the ground-state single-particle Green's function
$G(r) = \langle c_{i,\uparrow}^\dagger c_{j,\uparrow} \rangle + \langle c_{i,\downarrow}^\dagger c_{j,\downarrow} \rangle$
and the spin correlation 
$F(r) = \langle \mathbf{S}_i \cdot \mathbf{S}_j \rangle$ in the FFLO superconductive regime, 
where $r = |i - j|$ the denotes the distance along the $x$-direction.
Fig.~\ref{FigS5}(a) illustrates that $|G(r)|$ decays exponentially, indicating a finite single-particle gap.
In contrast, $|F(r)|$ in Fig.~\ref{FigS5}(b) exhibits power-law decaying behavior, suggesting the absence of spin gap
under Zeeman fields.

\section{Spin and charge density waves in real space}
\begin{figure}[htbp]
\includegraphics[width=0.7\textwidth]{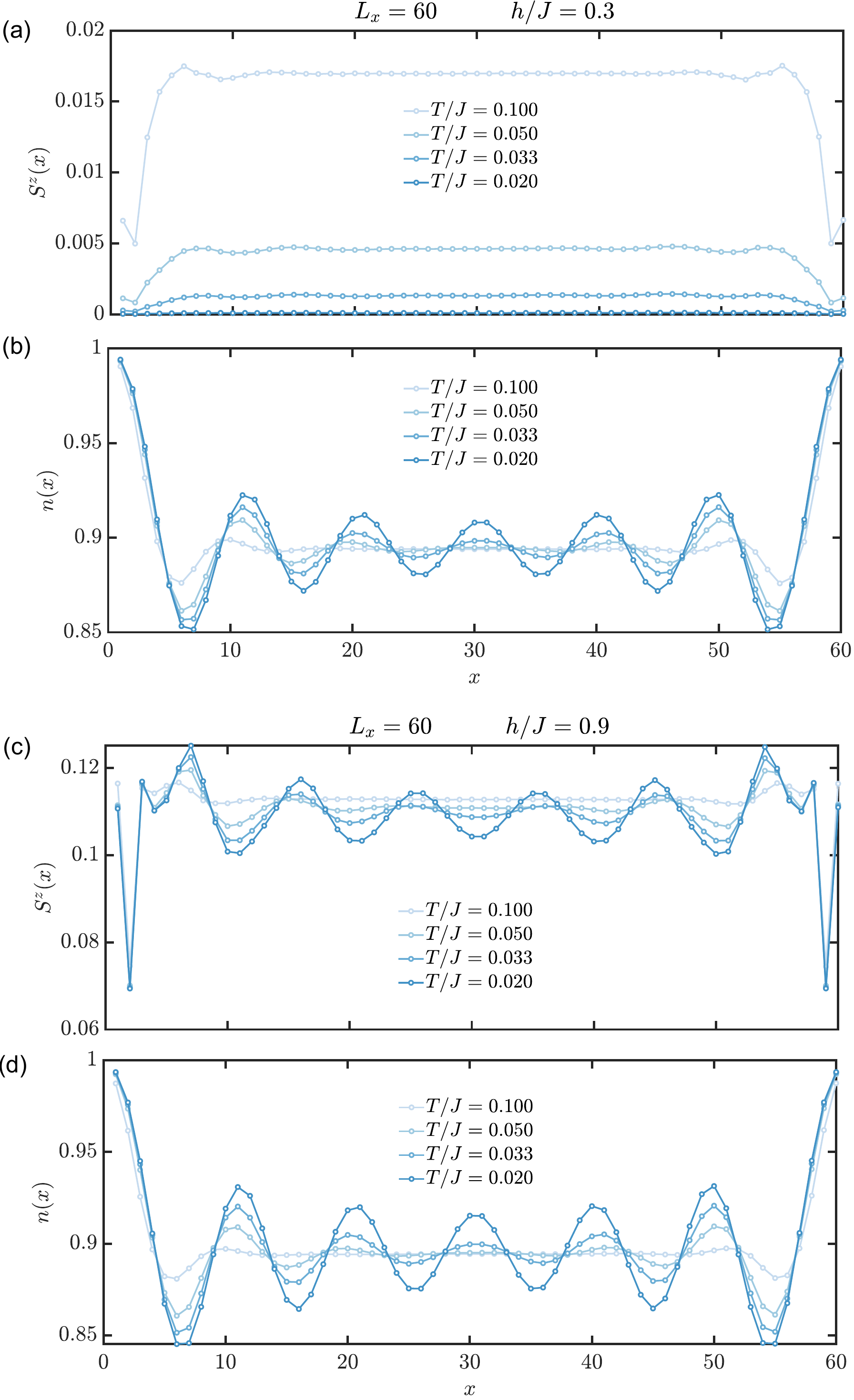}
\caption{Rung averaged charge density $n(x)$ and spin density $S^z(x)$ in the real space at several temperatures.
(a)(b) are taken from the $d$-SC + CDW (2e-SS1) phase at $h/J = 0.3$. 
(c)(d) correspond to the FFLO + SDW/CDW (2e-SS2) phase at $h/J = 0.9$.}
\label{FigS6}
\end{figure}

To further clarify the charge and spin density waves, we show in Fig.~\ref{FigS6} the rung averaged charge density 
$n(x)$ and spin density $S^z(x)$ in real space at various temperatures.
The calculated system is a $L_x = 60$, $L_y = 2$ ladder with doping $\delta \simeq 0.1$, equivalently 12 holes in 
120 lattice sites.
We take $h/J = 0.3$ in Figs~\ref{FigS6}(a, b), corresponding to the $d$-SC + CDW (2e-SS1) phase. 
The charge density profile emerges a pronounced density wave with 6 dips,
indicating two holes within each stripe and forms a ``filled-stripe'' pattern.
The spin density profile exhibits a uniform distribution. As temperature decreases, the magnetization $m$ 
converges to zero, consistent with the zero-magnetization ground state.
In Figs~\ref{FigS6}(c, d), we take $h/J = 0.9$, corresponding to the FFLO + SDW/CDW (2e-SS2) phase. 
As is discussed in the main text, the system exhibits a field-induced spin density wave, while the intertwined
charge density order gets suppressed by the Zeeman field.

\section{Results for the $W=4$ cylinders}
Going beyond the two-leg ladder case discussed in the main text, we further study the $t$-$t'$-$J$ model on 
width-4 cylinders and show the primary DMRG results in this section.
Previous researches~\cite{LE1LE2} have confirmed the Luther-Emery superconductive ground state 
of the width-4 $t$-$t'$-$J$ model in the absence of magnetic field. 
In contrast to the ordinary $d$-wave pattern that can exist within the 2D limit when $t'/t>0$, 
the geometry of a width-4 cylinder results in an artificial ``plaquette" $d$-wave superconductivity when $t'/t<0$~\cite{Plaq_d_wave}

\begin{figure*}[tbp]
\includegraphics[width=1.0\textwidth]{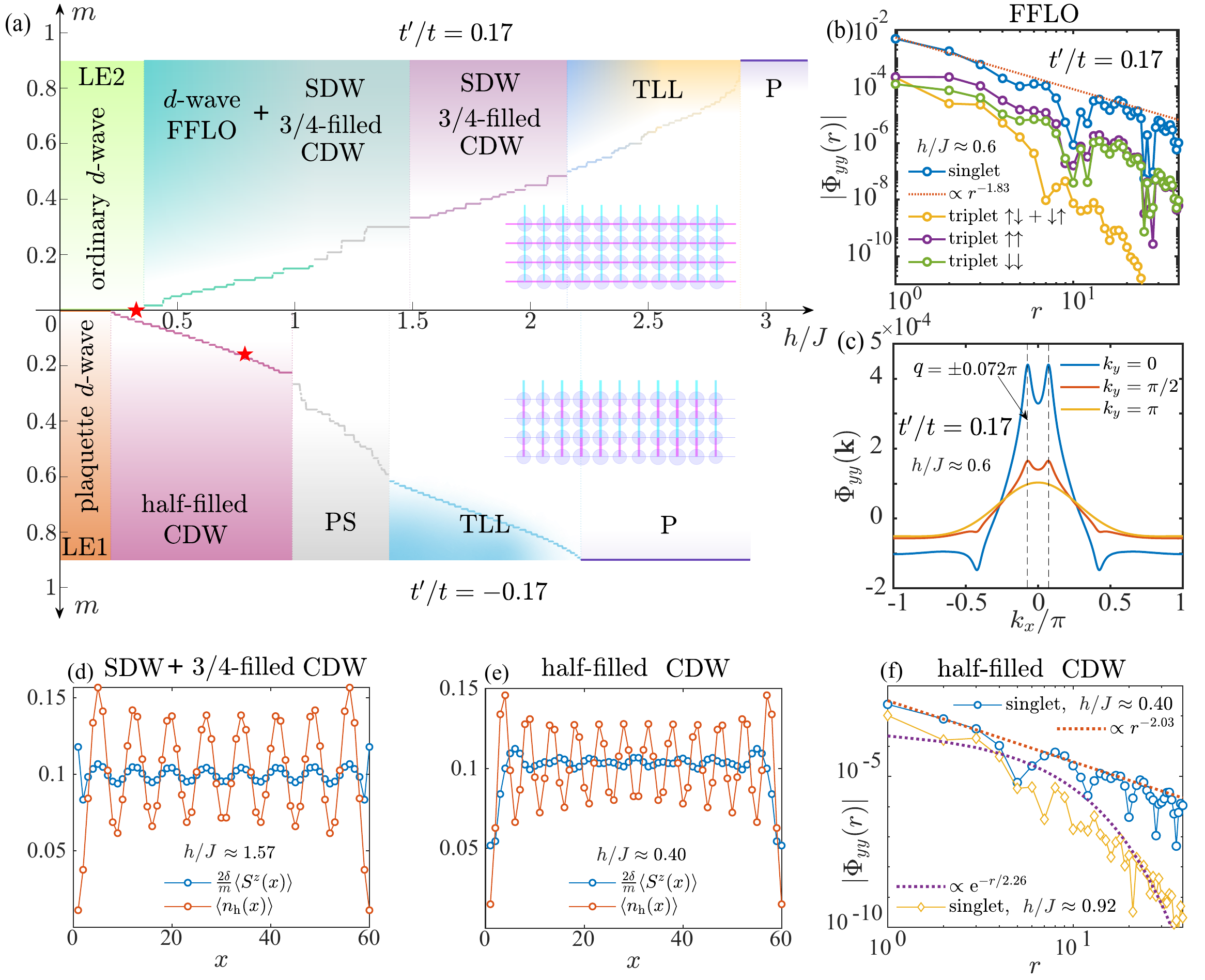}
\caption{(a) Magnetization curves for $t'/t = 0.17$ (top) and $t' / t = -0.17$ (bottom). The superconducting ground states are 
Luther Emery liquids (LE), exhibiting ordinary $d$-wave and ``plaquette" $d$-wave, respectively. The corresponding pairing pattern are illustrated in the insets.
The Pauli limits in both systems are marked by red stars.
Notably, we identify a $d$-wave FFLO phase in the $t'/t > 0$ regime, with coexisting SDW and CDW. 
The $d$-wave FFLO phase is suppressed by Zeeman field $h/J \gtrsim 1.5$, while the 
SDW and 3/4-filled CDW persist till $h/J \approx 2.1$.
However, the FFLO superconductivity is absent in the $t'/t < 0$ regime, where the half-filled CDW emerges once the 
spin gap is closed.
After the suppression of CDW, the systems move into Tomonaga-Luttinger liquid phases (TLL). 
There is also a phase separation (PS) between CDW and TLL in $t' /t < 0$ case.
The polarized state where spin exchange vanishes is marked by ``P".
(b) Pairing correlations in the $d$-wave FFLO phase ($t' / t = 0.17, h \approx 0.6 J$). We show $|\Phi_{yy}|$ for 
both singlet and three triplet channels. The dominant singlet pairing decays as $r^{-1.83}$, ensuring a 
quasi-long-range SC order in the ground state.
(c) Fourier transform of singlet $yy$ pairing correlations with non-zero peaks, indicating a finite-momentum pairing.
(d) Spin and charge density profiles in the SDW + 3/4-filled CDW phase ($t'/t = 0.17, h/J \approx 1.57$).
(e) Spin and charge density profiles in the half-filled CDW phase ($t'/t = -0.17, h/J \approx 0.4$).
(f) Pairing correlations in the half-filled CDW phase, ensuring an absence of superconductivity in the $t'/t = -0.17, h/J \gtrsim 0.4$ regions.
}
\label{FigS3}
\end{figure*}

We choose the same model parameters $t=3, J=1, t'=\pm0.51$ as the main text, fix the hole doping $\delta = 0.1$, 
and perform DMRG calculations up to $7000$ U(1)$_{\text{charge}} \times$ U(1)$_{\text{spin}}$ states to obtain the 
well-converged ground state of a $L_x = 60, L_y = 4$ cylinder.
Zeeman effect is introduced through scanning all the spin quantum numbers.
The obtained magnetization curves are shown in Fig.~\ref{FigS3}(a).
Consistent with the ladder case, we found a $d$-wave FFLO superconductive regime 
in the $t'/t > 0$ region. We check the $yy$ pairing correlations in both singlet and 
triplet channels in Fig.~\ref{FigS3}(b), where the singlet channel exhibits power-law decaying behavior 
$|\Phi_{yy}|(r) \propto r^{-1.83}$.
The triplet channels are less prominent and even decay exponentially.
We perform the Fourier transform and show the $\Phi_{yy}(\mathbf{k})$ in Fig.~\ref{FigS3}(c).
The finite-momentum $q = 0.072 \pi$ pairing indicates the emergence of FFLO phase.
Under a larger Zeeman field $h/J \gtrsim 1$, we further identify the coexistence of the FFLO superconductivity with the 
spin/charge density wave orders, which is consistent with the charge-2e supersolid mentioned in the main text.
The FFLO superconductivity is suppressed at $h/J \gtrsim 1.5$, 
above which the spin density wave and 3/4-filled charge density wave become dominant in the ground state (cf. Fig.~\ref{FigS3}(d)).
In the $t'/t < 0$ region, the ``plaquette" $d$-wave superconductivity is suppressed by the 
Zeeman field (cf. Fig.~\ref{FigS3}(f)) after the closure of spin gap 
and replaced by a half-filled charge density wave (cf. Fig.~\ref{FigS3}(e)). 
Both $t'/t >0$ and $t'/t <0$ systems evolve into Tomonaga-Luttinger liquid phases (TLL) under high Zeeman fields and finally polarize (P).

Comparing the results of 4-leg cylinders and 2-leg ladders, we could expect the emergence of FFLO superconductivity,
as well as the charge-2e supersolidity, in wider $t$-$t'$-$J$ cylindrical systems
and even in 2D limit under Zeeman field for $t'/t > 0$.

\end{document}